\documentclass[12pt]{iopart}
\pdfoutput=1

\expandafter\let\csname equation*\endcsname=\relax
\expandafter\let\csname endequation*\endcsname=\relax
\usepackage{color}
\usepackage{amsmath}
\usepackage{amssymb}
\usepackage{enumerate}
\usepackage{graphicx}
\usepackage{amsfonts}

\newcommand{\nn}{\nonumber}

\newcommand{\bea}{\begin{eqnarray}}
\newcommand{\eea}{\end{eqnarray}}
\newcommand{\beq}{\begin{equation}}
\newcommand{\eeq}{\end{equation}}

\def\XXint#1#2#3{{\setbox0=\hbox{$#1{#2#3}{\int}$}
 \vcenter{\hbox{$#2#3$}}\kern-.5\wd0}}

\usepackage[utf8]{inputenc}
\usepackage{amsmath}
\usepackage{hyperref}
\usepackage{graphicx}
\usepackage{amsfonts}
\usepackage{amsthm}
\usepackage{cases}
\usepackage{bm}
\usepackage{amssymb}

\usepackage{color}
\definecolor{Blue}{rgb}{0.00, 0.00, 1.00}
\definecolor{Red}{rgb}{1.00, 0.00, 0.00}

\hypersetup{
    colorlinks=true,       
    linkcolor=red,          
    citecolor=blue,        
    filecolor=magenta,      
    urlcolor=cyan           
}

\newcommand{\be}{\begin{equation}}
\newcommand{\ee}{\end{equation}}

\newcommand{\beqn}{\begin{eqnarray}}
\newcommand{\eeqn}{\end{eqnarray}}

\newcommand{\moy}[1]{\ensuremath{\left\langle #1 \right\rangle}}

\newcommand{\pFq}[5]{{}_{#1}\mathrm{F}_{#2} \left( \begin{array}{c} #3
\\ #4 \end{array} ; #5 \right)}

\makeatletter
\renewcommand\@appendixstar{\@@par
 \ifnumbysec
 \@addtoreset{table}{section}
 \@addtoreset{figure}{section}\fi
 \setcounter{section}{0}
 \setcounter{subsection}{0}
 \setcounter{subsubsection}{0}
 \setcounter{equation}{0}
 \setcounter{figure}{0}
 \setcounter{table}{0}
 \def\thesection{\Alph{section}} 
 \def\theequation{\ifnumbysec
      \Alph{section}.\arabic{equation}\else
      \Alph{section}\arabic{equation}\fi}
 \def\thetable{\ifnumbysec
      \Alph{section}\arabic{table}\else
      A\arabic{table}\fi}
 \def\thefigure{\ifnumbysec
      \Alph{section}\arabic{figure}\else
      A\arabic{figure}\fi}}
\makeatother
\begin{document}
\title[]{Counting equilibria in a  random non-gradient dynamics with heterogeneous relaxation rates}

\author{Bertrand Lacroix-A-Chez-Toine}
\address{Department of Mathematics, King’s College London, London
WC2R 2LS, United Kingdom}

\author{Yan V Fyodorov}
\address{Department of Mathematics, King’s College London, London
WC2R 2LS, United Kingdom}
\address{L.D. Landau Institute for Theoretical Physics, Semenova 1a, 142432 Chernogolovka, Russia}

\begin{abstract}
We consider a nonlinear autonomous random dynamical system of $N$ degrees of freedom coupled by Gaussian random interactions and characterized by a continuous spectrum $n_{\mu}(\lambda)$ of real positive relaxation rates. Using Kac-Rice formalism, the computation of annealed complexities (both of stable equilibria and of all types of equilibria) is reduced to evaluating the averages involving the modulus of the determinant of the random  Jacobian matrix. In the limit of large system $N\gg 1$ we derive exact analytical results for the complexities for  short-range correlated coupling fields,  extending results previously obtained for the ''homogeneous'' relaxation spectrum characterised by a single relaxation rate.
We show the emergence of a "topology trivialisation" transition from  a complex phase with exponentially many equilibria to a simple phase with a single equilibrium as the magnitude of the random field is decreased. Within the complex phase the complexity of stable equilibria undergoes an additional transition from a phase with exponentially small probability to find a single stable equilibrium to a phase with exponentially many stable equilibria as the fraction of gradient component of the field is increased. The behaviour of the complexity at the transition is found only to depend on the small $\lambda$ behaviour of the spectrum of relaxation rates $n_{\mu}(\lambda)$ and thus conjectured to be universal. We also provide some insights into a counting problem motivated by a paper of Spivak and Zyuzin of 2004 about wave scattering in a disordered nonlinear medium.
\end{abstract}
\vspace{2pc}

\maketitle

\section{Introduction and definition of the model}

Quantifying the number and characterizing the stability of dynamic equilibria of a large complex system describing
 the time evolution of $N$ interacting degrees of freedom in the form
\be
\partial_t x_i=-\mu_i\, x_i+f_i({\bf x})\;,\;\;{\bf x}=\{x_1,\cdots,x_N\}\;,\;\;i=1,\;\cdots,\;N\;,\label{evol_xi}
\ee
is a ubiquitous problem motivated by numerous applications in fields ranging from ecology \cite{M72} to economics \cite{HM11,FS13,MB19}. The parameters $\mu_i$ in Eq.(\ref{evol_xi}) control
the typical relaxation rates for individual degrees of freedom $x_i,\, i=1,\ldots,N$ whereas the fields $f_i({\bf x})$ provide their interaction. Trying to get an understanding of a generic rather than system-specific behaviour, it is natural to consider the interactions $f_i({\bf x})$ to be random functions.  The simplest question for this type of problems can be formulated as characterizing the total number of equilibria ${\cal N}_{\rm tot}$  which for the equations (\ref{evol_xi}) amounts to counting real solutions of the system of $N$ nonlinear equations

\be
\mu_i\, x_i=f_i({\bf x})\;,\;\;i=1,\;\cdots,\;N\;.\label{eqvil_xi}
\ee

 The nonlinear system Eq.(\ref{eqvil_xi}) may have multiple solutions whose
number and locations depend on the realization of the random fields $f_i$ and their numbering and classification by instability index (the number of unstable directions) is one of the most natural questions to be addressed before considering any more detailed characteristics.

Early line of research along this direction which started about 50 years ago with the seminal work \cite{M72} has been mainly concentrating on the simplest linear incarnation of the model: $f_i({\bf x})=\sum_{ij}J_{ij}x_j$ assuming homogeneous relaxation rates $\mu_i=\mu, \forall i$.  Such studies thus necessarily addressed stability of a single chosen equilibrium, with the local Jacobian $\frac{\partial f_i}{\partial x_j}$ replaced by a random matrix $J_{ij}$, see \cite{AT15} for a review, and \cite{MM21} for recent developments addressing effects of inhomogeneity in $\mu_i$. A promotion of these ideas to a nonlinear setting has been achieved only relatively recently \cite{WT13,FK16,F2016,BAFK21,FFI21,Ipsen}.

Most detailed investigations of the ensuing structures of equilibria in nonlinear systems  have been mainly performed
in a special case of gradient descent flows, characterized by
the existence of a potential function $V({\bf x})$ such that $f_i=-\frac{\partial V}{\partial x_i}$. In
this case the dynamical system  Eq.(\ref{evol_xi}) can be rewritten in the form $\partial_t{\bf x}=-\nabla L$, with
the (Lyapunov) function $L({\bf x})=\frac{1}{2}\sum_i\mu_i x_i^2+V({\bf x})$
describing the so called ''effective landscape''.
The associated ''random landscape paradigm'' originated in the theory of disordered systems such as  spin glasses, see \cite{CCrev} for an accessible introduction and \cite{Urb1,Urb2} for more recent examples of some models of this type., gradually became popular beyond the original setting finding numerous applications in such diverse fields as cosmology \cite{cosmol1,cosmol2}, machine learning via deep neural networks \cite{Choromanska_et_al,Bask1}, and large-size inference problems in statistics \cite{BMMN19,RABC20,FT20,MBAB20}. The dynamical equilibria in that case are associated with minima, maxima and saddle points on the corresponding landscape.  The quantities attracting a lot of interest in this context are
the so-called ''complexities'', i.e. the rates of exponential growth of the associated counting functions with parameter $N$.
Note that random local Jacobians in such cases are simply related to the local Hessians by $J_{ij}=\mu_i\delta_{ij}+\frac{\partial ^2 V}{\partial x_i \partial x_j}$ and hence are necessarily symmetric.
The problem of counting and classification of  stationary points in random energy landscapes was first addressed in the spin glass literature, see e.g. \cite{CGP98,CGG99} and references therein.  By using powerful methods like the replica trick those early works provided important insights into the structure of the landscapes, but some aspects of the calculations relied upon approximations which remained heuristic.

 The first fully controllable (and eventually mathematically rigorous) approach to the equilibria counting problem for the gradient flow variant of Eq.(\ref{evol_xi}) with homogeneous relaxation rates $\mu_i=\mu$ was proposed in \cite{F04} and further developed in  \cite{BD07,FW07,FN12,GK21}.  The approach started from the so-called Kac-Rice formula and heavily used the methods and results borrowed from the theory of large random  matrices (RMT), see \cite{F15} for a pedagogical introduction. In that framework one was able to provide explicit expressions for the so called ''annealed complexities'' given by the logarithm of the mean number of equilibria of a given instability index. One of the important insights stemming from that calculation was the identification of the so-called '' landscape topology trivialization'' transition occurring when the parameter $\mu$ exceeded a critical value $\mu_c$ set by the variance of the local Hessian.
 Such $\mu_c$ is exactly the point where the so called replica symmetry breaking mechanism ceases to be operative \cite{FW07} reflecting the change in the nature of the landscape from supporting exponentially many equilibria to a single equilibrium. This transition is further accompanied by the change in the Hessian spectrum at the global minimum of the associated landscape \cite{FLD18}.

 In a parallel independent development starting from the papers \cite{ABAC13a,ABAC13b} a very similar approach was suggested for counting stationary points of any index  in another class of random  potential landscapes, one characterizing the so-called spherical spin glasses. The ''topology trivialization'' transition in that setting manifests itself in vanishing complexity  with increasing the applied magnetic field \cite{F15,FLD14,BCNS21}.   In recent years the RMT methods have been successfully refined in the spherical case to achieve a rigorous control of higher moments of the counting function,  and eventually the typical (or ''quenched'') values of the associated complexities \cite{S17,SS17,S18,AG20,S21,SS21}. In parallel, much insight along similar lines has been obtained in the physics literature \cite{RABC20,R2020}.

Recently, the framework of the ''landscape topology trivialization transition'' phenomena has been essentially extended to deal with a physical problem of long-standing interest - the depinning transition of elastic manifolds in a random potential \cite{FLDRT18,FLD20_2}. Developing that line of research further the recent paper \cite{BABM21a} revealed its intimate connection to the equilibria counting problem for gradient flows Eq.(\ref{evol_xi})  with inhomogeneous rates $\mu_i$ characterized by a certain limiting density
\be
n_{\mu}(\lambda)=\lim_{N\to \infty}\frac{1}{N}\sum_{k=1}^N \delta(\mu_k-\lambda)\;.\label{denrate}
\ee
Annealed complexities in that case have been then rigorously and elegantly computed for a broad class of densities $n_{\mu}(\lambda)$ with a bounded support separated from zero by a gap. The paper \cite{BABM21a} used advanced RMT insights into the properties of expectations of random determinants obtained by the same authors in an accompanying paper \cite{BABM21b}. In a nutshell, they provided a rigorous proof
of the following asymptotic identity:
\begin{equation}\label{detselfaver}
\lim_{N\to \infty}\frac{1}{N}\ln{\moy{|\det H|}}=\lim_{N\to \infty}\frac{1}{N}\moy{\ln{|\det{H}|}} \equiv \int \rho(\lambda)\ln|\lambda|\,d\lambda
\end{equation}
for a broad class of $N\times N$ self-adjoint random matrices $H$, with  $ \rho(\lambda)$ standing for the associated limiting mean spectral density of real eigenvalues of $H$. Here and henceforth the angular brackets stand for the expectation/mean with respect to all relevant random variables.

The goal of the present work is to consider the problem of (annealed) equilibria counting with inhomogeneous relaxation beyond the assumptions of \cite{BABM21a},  namely
\begin{itemize}
\item (i) treating {\it non-gradient} flows
\item (ii) disposing with the assumption of the gapped density of relaxation rates.
\end{itemize}
In doing this we assume that in all cases under considerations the analogues of \eqref{detselfaver} (which is sometimes called the
property of strong self-averaging of the logarithm of mod-determinant) remain valid mutatis mutandis.
In particular,  for a class of non-selfadjoint random matrices $\rho(\lambda)$ needs to be replaced by the mean density $\rho(x,y)$
of complex eigenvalues of the corresponding matrices in the complex plane $z=x+iy$, with integration going over the plane.
 As this goes much beyond the proved theorems in several directions, our results should be considered as well-grounded conjectures
 from the point of view of rigorous mathematics.

 Although we will deal with such a problem in a considerable generality, our main  interest is in considering the case when the individual relaxation rates follow the power law scaling: $\mu_k=\mu(k/N)^{1/\eta}$ for $k=1,\ldots,N$ with exponent $\eta>0$. This choice then dictates that the corresponding limiting density $n_{\mu}(\lambda)$ for $N\to \infty$ is supported on an interval extending down to zero and vanishes as a power law at the origin, namely
\be
n_{\mu}(\lambda)=\frac{\eta}{\mu}\,\left(\frac{\lambda}{\mu}\right)^{\eta-1}\,\Theta(\lambda(\mu-\lambda))\label{pow_law_spec}\;,
\ee
where $\Theta(x)=1$ for $x>0$ and zero otherwise.

As it turns out, the above choice provides quite a rich phenomenology for behavior of landscape complexities at the topology trivialization transition. However our interest in the problem was not motivated by a purely academic curiosity, but rather prompted by a counting problem arising in physics of diffusive wave scattering in a nonlinear disordered media considered originally in the paper by B. Spivak and A. Zyuzin \cite{SZ04} which we briefly introduce below.

In the latter paper, the authors considered waves propagating in a three-dimensional sample of disordered non-linear medium of fixed linear extent $L$. Their starting point is the following nonlinear Schr\"odinger equation at a fixed wave energy $\epsilon$
\be\label{NLS}
\left[-\frac{\hbar^2}{2m}\Delta_{\bf r}-\epsilon +u({\bf r})+\beta n({\bf r})\right]\phi({\bf r})=0\;,
\ee
where $\phi({\bf r})$ is the wave function amplitude, $n(r)=|\phi({\bf r})|^2$ is the associated density, $\Delta_{\bf r}$ is the Laplacian operator, $u({\bf r})$ is a random potential characterising the disorder distributed inside the medium and $\beta$ controlling the strength of the non-linearity. While for a linear medium with $\beta=0$, and for fixed boundary conditions this equation has a unique solution for each realisation of the random potential, the authors argue that for a non-linear medium the number of solutions becomes eventually exponentially large  with the system size $L$. To show this they suggest to expand the wavefunction density $ n({\bf r})$ over a complete set of eigenstates  $n_i({\bf r})$ that satisfy the associated classical diffusion equation:
\be
n({\bf r})\propto \sum_{i=1}^{\infty}\sqrt{E_i}\, u_i\,n_i({\bf r})\;,\;\;-\Delta_{\bf r}n_i({\bf r})=E_i\,n_i({\bf r})\;,\;\;\int d^3 {\bf r}\,n_i({\bf r})n_j({\bf r})=\delta_{ij}\;,\label{exp_dens}
\ee
where the random coefficients $u_i$ depend on the realisation of the random potential $u({\bf r})$ and the $E_i$'s are the eigenvalues of the Laplacian operator in this geometry.  In the simplest cubic geometry the eigenvalues are clearly labeled by integer triples $(n_x,n_y,n_z)$ and are given by
\be
E_{n_x,n_y,n_y}=\frac{\pi^2}{L^2}(n_x^2+n_y^2+n_z^2)\,.
\ee
Then the number ${\cal N}(E)$ of such eigenvalues which do not exceed the value $E$ can be estimated for large enough $E$ as
\begin{align}
{\cal N}(E)&=\sum_{n_x,n_y,nz=1}^{\infty}\Theta\left(E-\frac{\pi^2}{L^2}(n_x^2+n_y^2+n_z^2)\right)\\
&\approx 4\pi\int_0^{\infty}dn\,n^2\,\Theta\left(E-\frac{\pi^2}{L^2}n^2\right)=\frac{4}{3\pi^2}(E\,L^{2})^{3/2}\;.
\end{align}
Ordering these eigenvalues in increasing order $E_1\leq E_2\leq \cdots$, one therefore may assume that the energy of the $i^{th}$ eigenvalue scales as
\be
E_i\propto \frac{i^{2/3}}{L^2}\;,\;\;i\gg 1\;,
\ee
and the limiting density of the $N$ lowest eigenvalues is thus of power law type as given by \eqref{pow_law_spec},  with the particular choice of exponent $\eta=3/2$ and $\mu=E_N$.
As to the random coefficients $u_i$,  substituting the expansion (\ref{exp_dens}) back to the nonlinear Schr\"odinger equation (\ref{NLS}) lead the authors to conclude that those coefficients must satisfy a set of self-consistency equations
\be
i^{2/3}\,u_i=\gamma\, f_i({\bf u})\;,\;\;{\bf u}=\{u_1,\cdots,u_N\}\;,\;\;i=1,\cdots,N\;,\;\;\gamma=\frac{\epsilon_0}{\epsilon}\left(\frac{L}{l}\right)^{3/2}\;,\label{SZ_eq}
\ee
where $l$ is the mean free path, $\epsilon_0$ is an energy scale associated with the incoming wave and ${\bf f}$ is a random field that depends on the realisation of the random potential. At this point of our exposition we deviated from \cite{SZ04} in truncating the expansion formally at a large but finite value of $N$, with the aim to consider the $N\to \infty$ limit. Such procedure seems essential to have the problem mathematically well-defined and amenable to a  controlled analysis.

The properties of the random field ${\bf f}({\bf u})$ were studied in \cite{SZ04} by analysing the associated Feynman diagrammatics for propagating waves in a disordered sample and claimed to have Gaussian statistics with covariances
\begin{align}
\moy{f_i({\bf u}) f_j({\bf u})}&=\delta_{ij}\;,\;\;\moy{f_i({\bf u}) \partial_{u_j} f_k({\bf u})}=0\;,\label{SZ_cov_1}\\
\moy{\partial_{u_i}f_j({\bf u}) \partial_{u_k} f_l({\bf u})}&=C_{ijkl}(\epsilon)=\epsilon^{2/3}\frac{(j/l)^{1/3}+(l/j)^{1/3}}{2(|i-k|+|j-l|+\epsilon)^{2/3}}\;,\label{SZ_tensor}
\end{align}
where, when compared with the Eq. (12) of \cite{SZ04} we added a cut-off $\epsilon>0$ such that $C_{ijij}(\epsilon)=1$ takes a finite value. This again ensures that the problem is well-posed.

One then may see that the problem of counting the number of solution of the non-linear Schr\"odinger equation is then equivalent to counting the solutions of the set of $N$ randomly coupled Eqs. \eqref{SZ_eq}. It is therefore exactly in the form
of  Eq.(\ref{eqvil_xi}) with rates $\mu_i$ characterized by the power law density Eq.(\ref{pow_law_spec}) with $\eta=3/2$
and a special, highly nontrivial covariance structure of the random fields.
The authors of \cite{SZ04} attempted to estimate the number of solutions of the above equations for large $\gamma\gg 1$ using crude heuristic arguments: they estimated that for each $i^{2/3}<\gamma$ the number of solutions is multiplied by a factor $\gamma\,i^{-2/3}$ and by a factor $1$ for $i^{2/3}>\gamma$, yielding
\be
\ln \moy{N_{\rm tot}}_{\rm SZ}\approx\sum_{i=1}^{\infty} \ln\max\left(1,\gamma\, i^{-2/3}\right)\approx \frac{2}{3}\gamma^{3/2}\;,\;\;\gamma\gg 1\;.\label{SZ_esti}
\ee
Although their main qualitative conclusion of the exponentially large number of solutions turns out to be
correct, the proposed estimate of the complexity rate proves to be far off the value following from the well-controlled analysis (see Fig. \ref{SZ_vs_KR} below). Putting the required calculation for this intriguing problem on the firm ground of Kac-Rice formalism was one of the main motivations of writing our paper in the present form.

\subsection{Definition of the model}

In this paper, we aim to study the expression of the annealed total complexity and annealed complexity of stable equilibria
for the autonomous dynamical system (\ref{evol_xi}) with a general spectrum of relaxation rates $\mu_i$
specified via a continuous density (\ref{denrate}).

The starting point in these calculations is the Kac-Rice formulae for the mean total number of solutions/stable solutions of the system of equations (\ref{eqvil_xi}), see \cite{F15,Maillard2020} for an informal introduction.
  They can be conveniently expressed in terms of the ensemble-averaged modulus of the determinant of the random Jacobian matrix $J$ with elements $J_{ij}({\bf x})=\delta_{ij}\mu_i-\partial_{x_i}f_j$
\begin{align}\label{KacRice}
\moy{{\cal N}_{\rm tot}}&=\int d^N {\bf x}\, \left\langle{\prod_{k=1}^N\delta\left(f_k({\bf x})-\mu_k\right)|\det(J({\bf x}))|}\right\rangle\;,\\
\moy{{\cal N}_{\rm st}}&=\int d^N {\bf x}\, \moy{\prod_{k=1}^N\delta\left(f_k({\bf x})-\mu_k\right)|\det(J({\bf x}))|\chi(J({\bf x}))}\;,
\end{align}
Here the expectation is taken over the random fields $f_i$  and $\chi(A)$ is an indicator function, equal to one only if all the eigenvalues of $A$ have positive real part and zero otherwise.
These equations providing the mathematically rigorous way of computing the number of solutions, the remaining task amounts to evaluating the expectations and subsequently extracting the leading exponential asymptotic behaviour as $N\to \infty$.  The standard choice which allows to make progress towards completing this programme is to consider $f_i({\bf x})$ to be mean-zero Gaussian random fields.  Motivated both by the issue of analytic tractability and guided by our main example of interest specified in (\ref{SZ_cov_1})-(\ref{SZ_tensor}) we will concentrate on covariance structure of the field $f_i({\bf x})$ and its derivatives $\partial_{x_i}f_j({\bf x})$ at fixed position ${\bf x}$ in the form
\begin{align}
\moy{f_i({\bf x}) f_j({\bf x})}=\moy{f_i f_j}&=a\,\delta_{ij}\;,\label{cov_1}\\
\moy{f_i({\bf x})\partial_{x_j}f_k({\bf x})}=\moy{f_i X_{jk}}&=0\;,\label{cov_2}\\
\moy{\partial_{x_i} f_j({\bf x}) \partial_{x_k} f_l({\bf x})}=\moy{X_{ij} X_{kl}}&=\frac{c}{N^{2\psi}}\,C_{ijkl}\;,\label{cov_3}
\end{align}
where we have denoted $f_i\equiv f_i({\bf 0})$ and $X^f=\left.\partial_{x_i}f_j\right|_{{\bf x}={\bf 0}}$, with $a>0$ and the value of the exponent $\psi$  chosen such that
\be
0<\lim_{N\to \infty} \Tr(X^2)=\lim_{N\to \infty}\sum_{k,l=1}^{N}\moy{X_{kl}X_{lk}}=\lim_{N\to \infty}\frac{1}{N^{2\psi}}\sum_{k,l=1}^{N}C_{kllk}<\infty\;.
\ee
This choice of the random fields  ensures the (local) statistical independence of the Jacobian matrix from the random field ${\bf f}$. These properties can be exploited for the essential simplification in Kac-Rice formulae (\ref{KacRice}) reducing them after standard calculations, see e.g. \cite{F15,FLD20_2}{,  to the following form
\begin{align}
\moy{{\cal N}_{\rm tot}}&=\frac{\moy{|\det J({\bf 0})|}}{\det D^{\mu}}=\frac{\moy{|\det(D^{\mu}-X^f)|}}{\det(D^{\mu})}\;\;\;\label{N_tot_exp}\\
\moy{{\cal N}_{\rm st}}&=\frac{\moy{|\det J({\bf 0})|\chi(J({\bf 0}))}}{\det D^{\mu}}=\frac{\moy{|\det(D^{\mu}-X^f)|\chi(D^{\mu}-X^f)}}{\det(D^{\mu})}\;,\label{N_st_exp}\\
D_{ij}^{\mu}&=\mu_i \delta_{ij}\;,\;\;X_{ij}^f=\left.\partial_{x_i} f_j\right|_{{\bf x}={\bf 0}}\;,\;\;1\leq i,j\leq N\;.
\end{align}
Our main interest is in the limit $N\to \infty$ where well-defined characteristics of the system are the {\it annealed complexities}:
\begin{align}
\Xi_{\rm tot}=&\lim_{N\to \infty}\frac{1}{N}\ln\moy{{\cal N}_{\rm tot}}=\lim_{N\to \infty}\frac{1}{N}\left[\ln\moy{|\det(D^{\mu}-X^f)|}-\ln\det(D^{\mu})\right]\;,\label{xi_tot}\\
\Xi_{\rm st}=&\lim_{N\to \infty}\frac{1}{N}\ln\moy{{\cal N}_{\rm st}}\label{xi_st}\\
=&\lim_{N\to \infty}\frac{1}{N}\left[\ln\moy{|\det(D^{\mu}-X^f)|\chi(D^{\mu}-X^f)}-\ln\det(D^{\mu})\right]\;,\nn
\end{align}
that depend smoothly on the main control parameter $c$ which sets the magnitude of the random fields, as well as on other parameters of the model. Our main task is to study these dependencies.\\[0.5ex]

{\bf Remark 1:} The above choice generalizes the model considered in \cite{FK16,BAFK21}  by allowing both for the rates inhomogeneity and  a generic covariance tensor in (\ref{cov_1}-\ref{cov_3}),  as well as a general value of the exponent $\psi$.  Recall that in \cite{FK16,BAFK21} those quantities were chosen such that $\psi=1/2$ and
\be
C_{ijkl}\equiv C_{ijkl}(\tau)=\delta_{ik}\delta_{jl}+\tau (\delta_{il}\delta_{kj}+\delta_{ij}\delta_{kl})\;.\label{elliptic_tensor}
\ee
where $\tau$ is a measure of the ratio between the gradient and the solenoidal components of the field. The latter  choice  is natural to call the case of "zero range correlation". Tackling the case of a generic tensor $C$ analytically to the very end turns out to be very challenging, and we will only consider the zero-range covariance tensor \eqref{elliptic_tensor} in the remaining of this issue. We nevertheless can use exact Eqs. \eqref{xi_tot}-\eqref{xi_st} to evaluate numerically the complexities for a more general tensor. In particular, for the tensor defined in \eqref{SZ_tensor} as introduced in \cite{SZ04} we have computed numerically in Fig. \ref{SZ_vs_KR} the logarithm of the average total number of equilibria as a function of parameter $\gamma$. The result looks far off the heuristic estimate in Eq.\eqref{SZ_esti}.  Unfortunately, deriving an analytical expression describing the number of solutions in Spivak-Zyuzin model remains outstanding and represent a serious challenge. A deeper analytical study of this more generic covariance structure will be presented in a separate publication. The present paper is a first stepping stone aiming at putting this problem on a firmer theoretical ground.\\[0.5ex]

\vspace{1ex}

\begin{figure}
\centering
\includegraphics[width=0.8\textwidth]{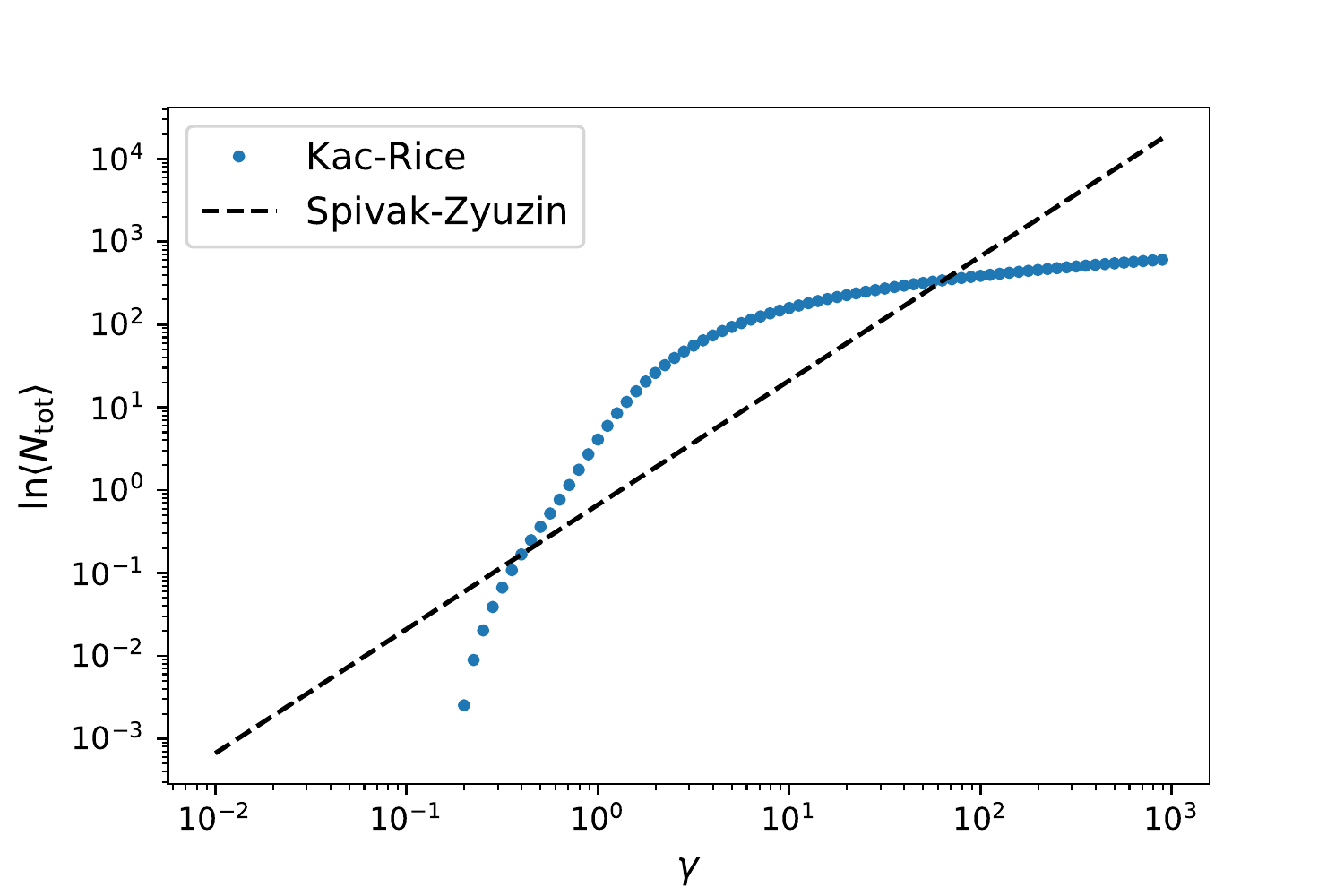}
\caption{The logarithm of the total number of equilibria for the model introduced in \cite{SZ04} as described by  Eq.\eqref{SZ_eq}
in the text plotted vs. parameter $\gamma$.  The blue dots correspond to a numerical evaluation based on the Kac-Rice formula: $\ln \moy{N_{\rm tot}}=\ln\moy{|\det(\mathbb{I}-X^f {D^{\mu}}^{-1})|}$ , with the choice $D_{ij}^{\mu}=i^{2/3}\delta_{ij}$ and the entries $X_{ij}^f=\gamma\partial_{u_i}f_j$ characterised by the covariance given in Eq.\eqref{SZ_tensor} for $\epsilon=1/4$.  The black dashed  line corresponds to the estimate Eq.\eqref{SZ_esti} as given in \cite{SZ04} and does not seem to reproduce correctly the large $\gamma$ behaviour of $\ln\moy{N_{\rm tot}}$. Note that due to the fast growth of the covariance tensor $C_{ijkl}$ with the truncation size $N$ as $N^4$
 we were only able to compute the complexity numerically by using truncations up to the size $N=10^2$.}\label{SZ_vs_KR}
\end{figure}

The rest of the paper is organised as follows. In section \ref{sec_res}, we summarise our main results and their implications. In section \ref{gen_met}, we provide a fully analytical expression for the annealed complexities when $X^f$ is of "zero range correlation" and relaxation rates are characterized by a generic continuous density $n_{\mu}(\lambda)$. We show in particular that varying the magnitude of the parameter $c$ one observes a landscape topology trivialization/detrivialization transition, separating the trivial phase where both $\Xi_{\rm tot}=0$ and $\Xi_{\rm st}=0$ for $c\leq c_t$ from the complex phase where $\Xi_{\rm tot}>0$ and $\Xi_{\rm st}\neq 0$ for $c>c_t$. Similarly, increasing the parameter $\tau$ controlling the ratio between the gradient and solenoidal components of the random field within the complex phase $c>c_t$, one observes in the complex phase a transition from $\Xi_{\rm st}<0$ for $\tau<\tau_0(c)$, i.e. an exponentially small probability to have any stable equilibrium, to  $\Xi_{\rm st}>0$ for $\tau>\tau_0(c)$, i.e. exponentially many stable equilibria. In the section \ref{power_law_sec} we then proceed to analyse in detail the special case  where the relaxation rates $\mu_k$'s are distributed according to a power law, implying
(\ref{pow_law_spec}). We compute in particular the behaviour of the complexities at the trivialization threshold separating the trivial and complex phases.
In section \ref{conclu}, we discuss the results and outline some perspectives for the future work. In the appendix \ref{AppA} we use two different methods to provide an explicit characterization of the density of complex eigenvalues for a real non-symmetric matrix $J=D^{\mu}+X$ where $D^{\mu}$ is a matrix with a known real spectrum and $X$ is a matrix from the real Elliptic Gaussian Ensemble. This generalizes the case of $X$ from Ginibre ensemble considered earlier in \cite{K96}, so this computation might be of interest on its own. Finally, in the appendix \ref{prop_elec_pot}, we derive some properties of the electrostatic potential associated with the density that allow us to derive our main results.

\ack
We would like to thank B. A. Khoruzhenko for explaining to us the details of the paper \cite{K96} and to P. Le Doussal for useful discussions. This research
was supported by the EPSRC Grant EP/V002473/1 {\it Random Hessians and Jacobians: theory and applications}.

\section{Summary of the main results}\label{sec_res}

\subsection{Random Matrix Results.} \label{RMTsec}
Our calculation of complexities in this case heavily relies on the following auxiliary results concerning complex spectra of large random non-selfadjoint matrices verified in the  Appendix (\ref{AppA}) by two different methods.\\[0.5ex]

Let us define the matrix
\be
J=D^{\mu}+X\;,
\ee
where $D_{ij}^{\mu}=\mu_i \delta_{ij}$ is a prescribed real diagonal matrix for which we know its limiting spectral density
\be
n_{\mu}(\lambda)=\lim_{N\to \infty}\frac{1}{N} \sum_{k=1}^{N}\delta(\mu_{k}-\lambda)\;,
\ee
 and $X$ is a matrix drawn from the real Gaussian Elliptic Ensemble with
\be
\moy{X_{ij}}=0\;,\;\;\moy{X_{ij}X_{kl}}=\frac{c}{N}(\delta_{ik}\delta_{jl}+\tau\,\delta_{il}\delta_{jk})\;.
\ee
Our aim is to obtain, in the limit $N\to \infty$, the closed-form expressions for the mean density of its complex eigenvalues
\be
\rho(z,\bar z)=\lim_{N\to \infty}\frac{1}{N}\sum_{k=1}^{N}\moy{\delta(z_{k}-z)\delta(\bar z_{k}-\bar z)}\;.
\ee

To this end we define for non-negative integers $l,m$ the following functions of two real variables $t\ge 0, q\in \mathbb{R}$:
\begin{equation}\label{3}
K_{l,m}(t,q)=\int_{\mathbb{R}} \frac{\lambda^l\,n_{\mu}(\lambda)\,d\lambda}{[t+(q-\lambda)^2]^m}
\end{equation}

Then one can demonstrate validity of the following\\[1ex]

{\bf Proposition}:
{\it As $N\to \infty$ the density
$$
\rho_N(z,\bar z):=\rho_N(x,y)=\frac{1}{N}\sum_i\delta(x-x_i)\delta(y-y_i)
$$
of complex eigenvalues $z_i=x_i+iy_i$ of the  matrices $J=D^{\mu}+X$  is positive only inside the domain with the boundary curve given by
 \begin{equation}\label{4}
K_{0,1}\left(t=\frac{y^2}{(1-\tau)^2},q(x)\right)=1
\end{equation}
where the function $q(x)$ (and its counterpart $t(x)>0$) are solutions of the system:
\begin{equation}\label{5}
  K_{0,1}(t,q)=1, \quad  \left(1+\tau\right)q-x=\tau  K_{1,1}(t,q)
  \end{equation}

Inside that domain the mean density tends to the function $\rho_{\infty}(x)=\lim_{N\to \infty}\rho_N(x,y)$ which is independent of $y$
and is given explicitly by:
 \begin{equation}\label{6}
\rho_{\infty}(x)=\frac{1}{\pi}\frac{1+D(x)}{1-\tau^2+2\tau D(x)}, \quad D(x)=L_{2,2}(x)-\frac{L^2_{1,2}(x)}{L_{0,2}(x)}
\end{equation}
where we defined $L_{l,m}(x):=K_{l,m}\left(t(x),q(x)\right)$.}

{\bf Note:} For $\tau=0$ our formulae are equivalent to those in \cite{K96}.\\[1ex]
{\bf Trivial Example:} Let $n_\mu(\lambda)=\delta(\lambda)$, implying $K_{0,m}=\frac{1}{(t+q^2)^m}$ and  $K_{l>0,m}(t,q)=0$.
Hence $q(x)=\frac{x}{1+\tau}$ and the boundary curve from (\ref{4}) is given by:
\[
\frac{y^2}{(1-\tau)^2}+\frac{x^2}{(1+\tau)^2}=1
\]
which is the ellipse with area $\pi(1-\tau)(1+\tau)=\pi(1-\tau^2)$. As $D(x)=0$,  the density $\rho_{\infty}(x)$ inside the ellipse is constant and is exactly equal to the inverse of the area - elliptic law.

\vspace{0.5cm}

\subsection{Annealed Complexities.}

We present below the full analytical expressions for both types of annealed complexities in the particular case where the Jacobian matrix only has short-range correlations, i.e. $\psi=1/2$ and the covariance tensor $C_{ijkl}\equiv C_{ijkl}(\tau)$ chosen according to Eq.(\ref{elliptic_tensor}). \\

\noindent {\it Total complexity}:
The total annealed complexity undergoes a transition from the simple (i.e. topologically trivial) phase for $c\leq c_t$ to the complex phase for $c>c_t$ where the total complexity $\Xi_{\rm tot}(c)$ and the critical threshold value $c_t$ are given respectively by
\be
\Xi_{\rm tot}(c)=\begin{cases}
0&\;,\;\;\displaystyle c\leq c_t=\left[\int\frac{d\lambda}{\lambda^2}\,n_{\mu}(\lambda)\right]^{-1}\;,\\
&\\
\displaystyle \int_{c_t}^c \frac{d\omega}{2\,\omega^2}\,\Upsilon^2(\omega)&\;,\;\;c> c_t\;.
\end{cases}
\ee
The function $\Upsilon(c)$ depends explicitly on the spectrum $n_{\mu}(\lambda)$ and should be found as the solution for $c>c_t$ of the equation
\be
\int d\lambda \,\frac{c\,n_{\mu}(\lambda)}{\lambda^2+\Upsilon^2(c)}=1\;.
\ee
In particular it is completely independent of the parameter $\tau$  controlling the ratio between the gradient and solenoidal components of the random field but only depends on its magnitude $c$. Additionally, given any rate density function $n_{\mu}(\lambda)$ for which the integrals
\be
I_p(\mu)=\int \frac{d\lambda}{\lambda^p}\,n_{\mu}(\lambda)\;,
\ee
are finite for $p\leq 4$, the total annealed complexity will vanish quadratically when approaching the threshold $c_t$ from above:
\be
\Xi_{\rm tot}(c)=\frac{I_2(\mu)^4}{4I_4(\mu)}(c-c_t)^2+o(c-c_t)^2\;,\;\;c\geq c_t\;.\label{quadratic}
\ee
The above results generalize a similar statement proven in the particular case $\tau=1$ of a gradient flow \cite{BABM21a}.

If however the density of rates $n_{\mu}(\lambda)$ is such that the integral $I_2(\mu)$ diverges, the complexity threshold $c_t\equiv 0$ and the corresponding system stays in the complex phase for any value of the coupling strength parameter $c>0$. The way in which the total annealed complexity vanishes as $c\to 0$ turns out to be controlled by an exponent that depends explicitly on the small $\lambda$ behaviour of $n_{\mu}(\lambda)$. Conversely, if the integral $I_2(\mu)$ is finite but $I_4(\mu)$ diverges, the
system enters the complex phase at a finite value $c=c_t>0$ but the total annealed complexity vanishes faster than quadratically at the threshold, with the critical behaviour again explicitly dependent on behaviour of $n_{\mu}(\lambda)$ as $\lambda\to 0$.

Such picture can be made very explicit for the power law scaled relaxation rates $\mu_k=\mu(k/N)^{1/\eta}$ characterized by the density (\ref{pow_law_spec}). In that case the small-scale density behaviour is controlled by the exponent $\eta$. The threshold value
$c_t$ remains at zero as long as $0<\eta\le 2$ and becomes positive for $\eta>2$. Correspondingly,  we find that the total complexity behaves close to the threshold $c_t$ as:
\be
\Xi_{\rm tot}(c)\approx\begin{cases}
\displaystyle \frac{2-\eta}{2\eta}\left(\frac{\eta \pi}{2 \sin\left(\frac{\eta \pi}{2}\right)}\right)^{\frac{2}{2-\eta}}\,\left(\frac{c}{\mu^2}\right)^{\frac{\eta}{2-\eta}}&\;,\;\;0<\eta<2\;,\\
&\\
\displaystyle \frac{1}{2}e^{-\mu^2/c}&\;,\;\;\eta=2\;,\\
&\\
\displaystyle \frac{\eta}{2(\eta-2)}\left[-\frac{2\eta\sin\left(\frac{\eta\pi}{2}\right)}{\pi (\eta-2)^2}\right]^{\frac{2}{\eta-2}}\left(\frac{c-c_t}{\mu^2}\right)^{\frac{\eta}{\eta-2}}&\;,\;\;2<\eta<4\;,\\
&\\
\displaystyle -\frac{2(c-c_t)^2}{\ln(2(c-c_t)/\mu^2)\mu^4} &\;,\;\;\eta=4\;,\\
&\\
\displaystyle \frac{\eta^3(\eta-4)}{4(\eta-2)^4}\left(\frac{c-c_t}{\mu^2}\right)^2&\;,\;\;\eta>4\;.
\end{cases}\label{xi_tot_res}
\ee
We see that for any $0<\eta\leq 4$ the critical behaviour of the total complexity $\Xi_{\rm tot}(c)$ is thus quite different from the quadratic behaviour in Eq. \eqref{quadratic}. Note also the essential singularity of the total complexity at the threshold for the special value $\eta=2$.  In general we expect the type of critical behaviour, such as the values of the critical  exponents  to be universal, i.e. independent of the details of $n_{\mu}(\lambda)$ apart from its behaviour on approaching the origin $\lambda\to 0$.\\

\noindent {\it Complexity of stable equilibria}: The annealed complexity of stable equilibria depends both on  the magnitude $c$ of the random field and the parameter $\tau$ controlling ratio of gradient components. It is obviously zero for $c<c_t$ as is the case for the total complexity and is given by
\be
\Xi_{\rm st}(c,\tau)=
\displaystyle \int_{c_t}^c \frac{\nu^2(\omega)}{2\omega^2}\,d\omega-\frac{(1-\tau)}{2c \tau}\nu^2(c), \quad c>c_t\;,
\ee
where the function $\nu(c)$ is independent of $\tau$ and satisfies
\be
\int d\lambda\,\frac{c\,n_{\mu}(\lambda)}{(\lambda+\nu(c))^2}=1\;.
\ee
For any rate density $n_{\mu}(\lambda)$ rendering the integral $I_3(\mu)=\int d\lambda \,\frac{n_{\mu}(\lambda)}{\lambda^3}$ finite, the complexity of stable equilibria vanishes at the threshold as
\be
\Xi_{\rm st}(c,\tau)=\left[\frac{c-c_t}{3\,c_t}-\frac{(1-\tau)}{\tau}\right]\frac{I_2^{5}(\mu)}{8\,\tau\,I_3(\mu)^2}(c-c_t)^2+o(c-c_t)^3\;,\;\;c\geq c_t\;.
\ee
This expression generalises the $\tau=1$ result in \cite{BABM21a}. Similarly to the total complexity, for densities $n_{\mu}(\lambda)$ rendering the integral $I_2(\mu)$ divergent  the annealed complexity of stable equilibria stays non-zero for any positive value $c>0$ of the random field. In that case as $c\to 0$ the complexity vanishes with an exponent that depends explicitly on the small $\lambda$ behaviour of $n_{\mu}(\lambda)$. Conversely, if the integral $I_2(\mu)$ is finite but $I_3(\mu)$ diverges, the complexity of stable equilibria vanishes at a finite threshold value $c=c_t=O(1)$ (identical to the critical value for the total complexity) with the exponent that depends explicitly on the small $\lambda$ behaviour of the rate density. Again, the explicit behaviour can be
found for the model with power law- scaled rates:
\be
\Xi_{\rm st}(c,\tau)\approx\begin{cases}
\displaystyle \frac{1}{2}\left(\frac{2}{\eta}-\frac{1}{\tau}\right)\left(\frac{\eta(1-\eta)}{\sin(\eta \pi)}\right)^{\frac{2}{2-\eta}}\left(\frac{c}{\mu^2}\right)^{\frac{\eta}{2-\eta}}&\;,\;\;0<\eta<2\;,\\
&\\
\displaystyle \frac{1}{2}\left(1-\frac{\mu^2(1-\tau)}{c\,\tau}\right)e^{-\mu^2/c-2}&\;,\;\;\eta=2\;,\\
&\\
\displaystyle \frac{\eta}{2(\eta-2)}\left(\frac{c-c_t}{\mu^2}-\frac{1-\tau}{\tau}\right)\left[\frac{\eta\sin(\eta\pi)}{(\eta-1)(\eta-2)^2\pi}\left(\frac{c-c_t}{\mu^2}\right)\right]^{\frac{2}{\eta-2}}&\;,\;\;2<\eta<3\;,\\
&\\
\displaystyle \left(\frac{c-c_t}{\mu^2}-\frac{1-\tau}{\tau}\right)\frac{27(c-c_t)^2}{8\ln^2((c-c_t)/\mu^2)\mu^4} &\;,\;\;\eta=3\;,\\
&\\
\displaystyle \frac{\eta^3(\eta-3)^2}{8(\eta-2)^5}\left(\frac{c-c_t}{3\,c_t}-\frac{1-\tau}{\tau}\right)\left(\frac{c-c_t}{\mu^2}\right)^2&\;,\;\;\eta>3\;.
\end{cases}\label{xi_stab_res}
\ee
We expect again this behaviour to be universal in the sense explained above.\\

{\bf Remark 2.}  As was first found in \cite{BAFK21}, in the complex phase
 the complexity of stable equilibria may experience an additional transition as a function of $\tau$. Namely, the complexity of stable equilibria is positive for $\tau\leq \tau_0(c)$, yielding exponentially many stable equilibria, while it is negative for $\tau>\tau_0(c)$, yielding an exponentially small probability to have any stable equilibrium. The expression of $\tau_0(c)$ reads
\be
\tau_0(c)=\frac{\nu^2(c)}{c\int_{c_t}^c \frac{\nu^2(\omega)}{\omega^2}\,d \omega+\nu^2(c)}\;,\;\;c\geq c_t\;.
\ee
A phase diagram with the behaviours of the total complexity and complexity of stable equilibria as a function of $c$ and $\tau$ is shown in Fig. \ref{phase_diag}.

\begin{figure}
\centering
\includegraphics[width=0.8\textwidth]{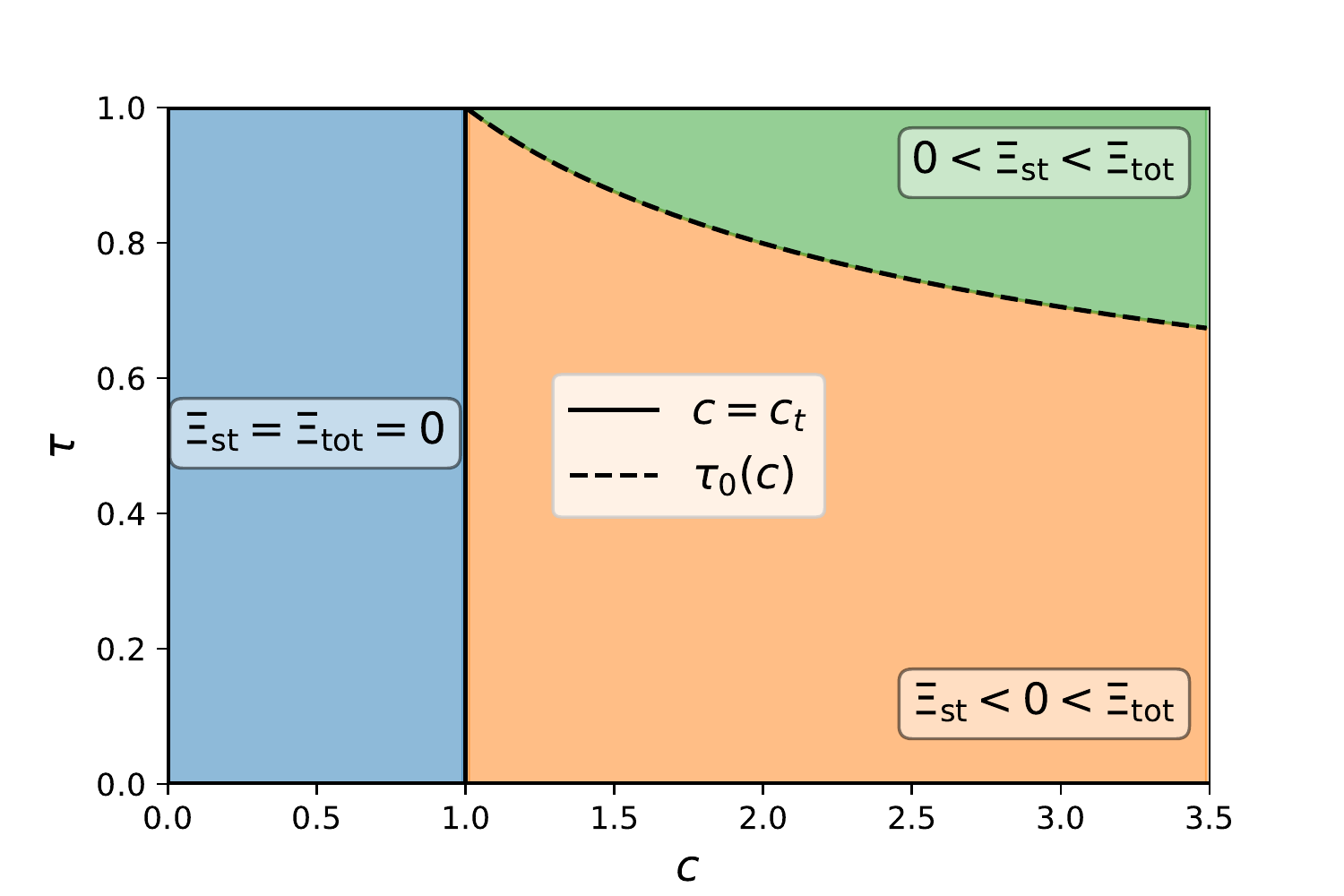}
\caption{Phase diagram showing the behaviour of the annealed total complexity $\Xi_{\rm tot}(c)$ and annealed complexity of stable equilibria $\Xi_{\rm st}(c,\tau)$ as a function of the magnitude of the random field $c$ and the fraction of gradient components $\tau$. In the trivial phase (in blue), i.e. for $c\leq c_t$ (marked by the continuous vertical line), both complexities are zero $\Xi_{\rm tot}(c\leq c_t)=\Xi_{\rm st}(c\leq c_t,\tau)=0$, while in the complex phase the total complexity is positive $\Xi_{\rm tot}(c>c_t)>0$. In the complex phase, the complexity of stable equilibria $\Xi_{\rm st}(c> c_t,\tau<\tau_0(c))<0$ is negative (in the orange region) below the line $\tau=\tau_0(c)$ (marked by the dashed black line) and positive above this line $\Xi_{\rm st}(c> c_t,\tau>\tau_0(c))>0$ (in the green region).}\label{phase_diag}
\end{figure}

\section{Derivation of complexities for a generic zero-range correlated Gaussian random field }\label{gen_met}

In this section we aim to analyse the expressions for the total complexity and the complexity of stable equilibria given respectively in Eqs. \eqref{xi_tot} and \eqref{xi_st} in the case of a zero-range correlated Gaussian random field. The Jacobian matrix only has short-range correlations, i.e. $\psi=1/2$ and the covariance tensor chosen according to Eq.(\ref{elliptic_tensor}), and can be equivalently expressed as
\be
X^f=\sqrt{c}\left(X-\sqrt{\frac{\tau}{N}}\, \chi\,\mathbb{I}\right)\;,
\ee
where $\mathbb{I}$ is the identity matrix, $\chi\sim N(0,1)$ is a real Gaussian random variable, $X$ is drawn from the real Elliptic Gaussian Ensemble (rEGE):
\be\label{EA}
\moy{X_{ij}}=0\;,\;\;\moy{X_{ij}X_{kl}}=\frac{1}{N}\left(\delta_{ik}\delta_{jl}+\tau \delta_{il}\delta_{kj}\right)\;,
\ee
We also consider the relaxation rate density $n_{\mu}(\lambda)$ for $D^{\mu}$ to have a generic form. Our considerations then generalize the single rate model $n_{\mu}(\lambda)=\delta(\lambda-\mu)$ studied in \cite{FK16,BAFK21}.

\subsection{Total complexity}

Let us first consider the total complexity. The numerator in Eq. \eqref{N_tot_exp} reads for finite $N$
\begin{align}
\moy{|\det(D^{\mu}-X^f)|}=\sqrt{\frac{N}{2\pi\sqrt{c\,\tau}}}\int_{-\infty}^{\infty} d\xi\, e^{\displaystyle -\frac{N \xi^2}{c\,\tau}}\moy{\left|\det(D^{\mu}+\xi\mathbb{I}-\sqrt{c}X)\right|}_{rEGE}\;,\label{den_N_tot}
\end{align}
where the average is taken over rEGE matrices $X$ defined in (\ref{EA}). Assuming validity of the spectral rigidity arguments for rEGE
\cite{BAFK21} put for $\tau=1$ on the firm mathematical ground in the gapped and bounded spectrum case in \cite{BABM21a,BABM21b} we rewrite in the large $N$ limit:
\begin{align}
\moy{|\det(D^{\mu}+\xi\mathbb{I}-\sqrt{c}X)|}_{rEGE}&=\moy{e^{\Tr\ln|D^{\mu}+\xi\mathbb{I}-\sqrt{c}X|}}_{rEGE}\label{SA_rEGE}\\
&\asymp e^{\moy{\Tr\ln|D^{\mu}+\xi\mathbb{I}-\sqrt{c}X|}_{rEGE}}\;,\nn
\end{align}
where by $\asymp$ we mean asymptotic logarithmic equivalence as $N\to \infty$. We will refer to the last relation as the self-averaging hypothesis.

 For $\tau\neq 1$ the eigenvalues of $D^{\mu}-\sqrt{c}X$ are to be complex and have the well-known interpretation in terms of a gas of charged particles \cite{F10}. In particular, the mean spectral density $\rho(z,\bar z)$ of eigenvalues at the point with coordinates $x=(z+\bar z)/2,\,y=(z-\bar z)/2i$ in the complex plane is related to the electrostatic potential $\Phi(z,\bar z;c,\tau)$ by the Poisson equation:
\be
\rho(z,\bar z)=\frac{1}{N}\sum_{k=1}^N \moy{\delta(z-z_k)\delta(\bar z-\bar z_k)}_{rEGE}=\frac{2}{\pi}\partial_{z \bar z}^2 \Phi(z,\bar z;c,\tau)\;,
\ee
implying the relation
\be
\Phi(z,\bar z;c,\tau)=\frac{1}{2}\int d^2 w \, \rho(w,\bar w)\ln(z-w)(\bar z-\bar w)\;.
\ee
Introducing such electrostatic potential for the density of complex eigenvalues of the matrices $D^{\mu}-\sqrt{c}X$ and using the self-averaging hypothesis Eq. \eqref{SA_rEGE} allows to re-express Eq. \eqref{den_N_tot} as
\be\label{asy_tot}
\moy{|\det(D^{\mu}-X^f)|}\asymp \int_{-\infty}^{\infty} d\xi\, e^{\displaystyle N\left[-\frac{\xi^2}{2\,c\,\tau}+\Phi(-\xi,-\xi;c,\tau)\right]}\;.
\ee
The total complexity, defined in Eq. \eqref{xi_tot}, can then be immediately extracted after evaluating the above integral in the limit $N\to \infty$ by the Laplace/saddle-point method. This gives
\begin{align}
\Xi_{\rm tot}(c,\tau)&=-\frac{\xi_*^2}{2\,c\,\tau}+\Phi(-\xi_*,-\xi_*;c,\tau)-\int d\lambda \, n_{\mu}(\lambda)\ln|\lambda|\;,\label{xi_tot_2}
\end{align}
where $\xi_*$ is the value of $\xi$ at the saddle-point satisfying
\begin{align}
\xi_*&=-\Gamma_{c,\tau}(-\xi_*)\;,\;\;\Gamma_{c,\tau}(x)=\left.c\,\tau\,\partial_x\Phi(z,\bar z;c,\tau)\right|_{z=\bar z=x}\;,\;\;z=x+i y\;.\label{sp_eq_2}
\end{align}

Obviously, further analysis of the total complexity hinges on availability of the spectral density $\rho(z,\bar z)$ and its associated electrostatic potential $\Phi(z,\bar z;c,\tau)$. In the appendix \ref{AppA} we provide two alternative derivations for the density $\rho(z,\bar z)$ in explicit form for a given arbitrary real spectrum $n_{\mu}(\lambda)$. In particular, we derive the following integral equations
\begin{align}
\Gamma_{c,\tau}(x)&=\int d\lambda \frac{c\,\tau\,n_{\mu}(\lambda)\,(x-\lambda-\Gamma_{c,\tau}(x))}{\displaystyle (x-\lambda-\Gamma_{c,\tau}(x))^2+\frac{y^2}{(1-\tau)^2}+\gamma_{c,\tau}^2(z,\bar z)}\;,\label{Gamma_eq}\\
1&=\int d\lambda \frac{c\,n_{\mu}(\lambda)}{\displaystyle(x-\lambda-\Gamma_{c,\tau}(x))^2+\frac{y^2}{(1-\tau)^2}+\gamma_{c,\tau}^2(z,\bar z)}\;,\label{gamma_eq}
\end{align}
holding for any positions $(z,\bar z)$ within the support of $\rho(z,\bar z)$.\\

\noindent {\bf Remark 3.1}. Within this support, the function $\gamma_{c,\tau}(z,\bar z)>0$ depends both on $x$ and $y$, while the function $\Gamma_{c,\tau}(x)$ turns out to be independent of $x$.

\noindent {\bf Remark 3.2}. Eq. \eqref{gamma_eq} loses its validity beyond the spectral support, where the function $\gamma_{c,\tau}(z,\bar z)=0$ while Eq. \eqref{Gamma_eq} remains valid albeit with $\Gamma_{c,\tau}(z,\bar z)$ depending both on $x$ and $y$. The boundary of the spectral support consists of points with coordinates $z_e(x_e)=x_e+ i y_e(x_e)$ satisfying both Eq. \eqref{gamma_eq} and $\gamma_{c,\tau}(z_e,\bar z_e)=0$. This yields
\begin{align}
\Gamma_{c,\tau}(x_e)&=\int d\lambda \frac{c\,\tau\,n_{\mu}(\lambda)\,(x_e-\lambda-\Gamma_{c,\tau}(x_e))}{\displaystyle(x_e-\lambda-\Gamma_{c,\tau}(x_e))^2+\frac{y_e^2(x_e)}{(1-\tau)^2}}\;,\\
1&=\int d\lambda \frac{c\,n_{\mu}(\lambda)}{\displaystyle(x_e-\lambda-\Gamma_{c,\tau}(x_e))^2+\frac{y_e^2(x_e)}{(1-\tau)^2}}\;.
\end{align}

\noindent {\bf Remark 3.3}. In the appendix \ref{prop_elec_pot} we derive the following two useful identities:
\begin{align}
\partial_{c}\Phi(x,x;c,\tau)&=\frac{1}{2\,c^2}\left(\gamma_{c,\tau}^2(x,x)-\frac{\Gamma_{c,\tau}^2(x)}{\tau}\right)\;,\label{phi_c}\\
\partial_{\tau}\Phi(x,x;c,\tau)&=-\frac{\Gamma_{c,\tau}^2(x)}{2\,c\,\tau^2}\;.\label{phi_tau}
\end{align}

With this in hand, we can take the derivative of Eq. \eqref{xi_tot_2} with respect to $\tau$, and exploiting Eq. \eqref{phi_tau} together with the saddle-point equation $\xi_*=-\Gamma_{c,\tau}(-\xi_*)$ in \eqref{sp_eq_2} can show that
\be
\partial_{\tau} \Xi_{\rm tot}(c,\tau)=\left[-\frac{\xi_*}{c\,\tau}+\left.\partial_{x}\Phi\right|_{z=\bar z=-\xi_*}\right]\partial_{\tau}\xi_*+\frac{\xi_*^2}{2\,c\,\tau^2}+\partial_{\tau}\Phi(-\xi_*,-\xi_*;c,\tau)=0.
\ee
We thus conclude that the generic total complexity $\Xi_{\rm tot}(c,\tau)\equiv \Xi_{\rm tot}(c)$ is independent of the value of the non-potentiality  parameter $\tau$ ( for the special case $n_{\mu}(\lambda)=\delta(\lambda-\mu)$ this was first observed in \cite{FK16}).

On the other hand, taking the derivative of Eq. \eqref{xi_tot_2} with respect to parameter $c$ and using Eq. \eqref{phi_c} together with \eqref{sp_eq_2} one can obtain the following simple expression for the total complexity (cf.  $\partial_c \Xi_{\rm tot}$ recently derived in \cite{BABM21a} in the special case $\tau=1$) :
\begin{align}
\Xi_{\rm tot}(c)&=\int_0^c d\omega\,\left[\frac{\xi_*^2}{2\,\omega^2\,\tau}+\left.\partial_{c}\Phi(-\xi_*,-\xi_*;c,\tau)\right|_{c=\omega}\right]=\int_0^c \frac{d\omega}{2\,\omega^2}\,\Upsilon^2(\omega)\;,
\end{align}
where we have defined $\Upsilon(c):=\gamma(-\xi_*,-\xi_*)$ which satisfies
\be\label{Upsilon}
\Upsilon(c)=c\Upsilon(c)\int d\lambda \frac{n_{\mu}(\lambda)}{\lambda^2+\Upsilon^2(c)}\;,
\ee
and have further used that the complexity must vanish for $c\to 0$, hence \eqref{xi_tot_2} implies
\be
\lim_{c\to 0}\left[-\frac{\xi_*^2}{2\,c\,\tau}+\Phi(-\xi_*,-\xi_*;c,\tau)\right]=\int d\lambda\,n_{\mu}(\lambda)\ln|\lambda|\;.
\ee

Recall that $\Upsilon(c)$ must vanish as long as the (real) $-\xi_*$ is outside of the support of $\rho(z,\bar z)$. The latter condition
then amounts to $-\xi_*\leq \lambda_-$ with $\lambda_-$ being the leftmost point in the boundary $z_e$ of
 the eigenvalue support domain, where the boundary crosses the $x-$axis. Therefore it can be found from solving the system
 \eqref{Gamma_eq}-\eqref{gamma_eq} by setting $y_e=0$:
\begin{align}
\Gamma_{c,\tau}(\lambda_-)&=\int d\lambda \frac{c\,\tau\,n_{\mu}(\lambda)}{\lambda_- -\lambda-\Gamma_{c,\tau}(\lambda_-)}\;,\label{Gamma_l_-}\\
1&=\int d\lambda \frac{c\,n_{\mu}(\lambda)}{(\lambda_- -\lambda-\Gamma_{c,\tau}(\lambda_-))^2}\;.\label{gamma_l_-}
\end{align}
Using this one can check that $\Xi_{\rm tot}(c)=0$ for any $c\leq c_t$ such that $-\xi_*=\Gamma_{c_t,\tau}(-\xi_*)=\lambda_-$, that is
\be
c_t=I_2(\mu)^{-1}\;,\;\;I_2(\mu)=\int \frac{d\lambda}{\lambda^2}\,n_{\mu}(\lambda)\;.\label{c_t_eq}
\ee
We therefore arrive to essentially the same scenario for the total complexity in our non-potential model as was discussed in  \cite{BABM21a} in the gradient case $\tau=1$ , which is hardly surprising given that the total complexity $\Xi_{\rm tot}(c)$ is independent of $\tau$. The total complexity  for the zero-range correlation model is thus given generally as
\be
\Xi_{\rm tot}(c)=\begin{cases}
0&\;,\;\;c\leq c_t\;,\\
&\\
\displaystyle \int_{c_t}^c \frac{d\omega}{2\,\omega^2}\,\Upsilon^2(\omega)&\;,\;\;c> c_t\;,
\end{cases}\label{xi_tot_exp_sec}
\ee
where  $\Upsilon(c)$ for $c>c_t$ is the non-zero solution of \eqref{Upsilon}. For any spectrum $n_{\mu}(\lambda)$ such that both the integrals $I_2(\mu)$ and
\be
I_4(\mu)=\int \frac{d\lambda}{\lambda^4}\,n_{\mu}(\lambda)\;,
\ee
are well-defined and finite, one can Taylor-expand the equation (\ref{Upsilon}) for $c$ approaching $c_t$ from above, yielding
\be
1=\int d\lambda \frac{c\,n_{\mu}(\lambda)}{\lambda^2+\Upsilon^2(c)}=\frac{c}{c_t}-c_t I_4(\mu)\Upsilon^2(c)+o(c-c_t)^2\;.
\ee
Using this result, we find that the complexity vanishes quadratically as $c\to c_t$
\be
\Xi_{\rm tot}(c)=\frac{I_2(\mu)^4}{4I_4(\mu)}(c-c_t)^2+o(c-c_t)^2\;,\;\;c\geq c_t\;,\label{quad_van}
\ee
where we have used that $c_t=I_2(\mu)^{-1}$. The quadratic behaviour at the threshold and the associated pre-factor again match exactly the gradient case results \cite{BABM21a}.

\subsection{Complexity of stable equilibria}

Let us now consider the complexity of stable equilibria.  In what follows $\chi(A)$ stands for the indicator function equal to one only if all the eigenvalues of a matrix $A$ have positive real parts, and vanishing otherwise. Proceeding similarly to the case of the total complexity we first express the expectation value in the numerator of Eq. \eqref{N_st_exp} for a given matrix size $N$ as
\begin{align}
&\moy{|\det(D^{\mu}-X^f)|\chi(D^{\mu}-X^f)}\nn\\
&=\sqrt{\frac{N}{2\pi\sqrt{c\,\tau}}}\int_{-\infty}^{\infty} d\xi\, e^{\displaystyle -\frac{N \xi^2}{c\,\tau}}\moy{|\det(D^{\mu}+\xi\mathbb{I}-\sqrt{c}X)|\chi(D^{\mu}+\xi\mathbb{I}-\sqrt{c}X)}_{rEGE}\;,\label{den_N_min}
\end{align}
 Let $z_-$ be the (random) eigenvalue of $D^{\mu}-\sqrt{c}X$ with the smallest real part and denote $\lambda_{-}=\moy{\Re(z_-)}$.
  For estimating the asymptotic exponential growth rate of the right-hand side in \eqref{den_N_min} we follow the background discussion in \cite{BAFK21} and exploit the ideas of the large deviation theory in the context of random matrices.
   Along these lines one expects that for any $\xi>-\lambda_{-}$ one can write asymptotically
\be
|\det(D^{\mu}+\xi\mathbb{I}-\sqrt{c}X)|\chi(D^{\mu}+\xi\mathbb{I}-\sqrt{c}X)\approx |\det(D^{\mu}+\xi\mathbb{I}-\sqrt{c}X)|\;.
\ee
which as we know grows exponentially with $N$.  In contrast, for  $\xi<-\lambda_{-}$ the probability of the event $\chi(D^{\mu}+\xi\mathbb{I}-\sqrt{c}X)\ne 0$ is of the order $\exp{(- C\, N^2)}$
for some finite positive constant $C>0$ making the corresponding expectation under the $\xi-$integral in \eqref{den_N_min} asymptotically negligible. Recalling the definition of the electrostatic potential  we therefore can write for large enough $N$ (cf. \eqref{asy_tot}):
\be\label{asy_stab}
\moy{|\det(D^{\mu}-X^f)|\chi(D^{\mu}-X^f)}\asymp\int_{-\lambda_-}^{\infty} d\xi\, e^{\displaystyle N\left[-\frac{\xi^2}{2\,c\,\tau}+\Phi(-\xi,-\xi;c,\tau)\right]}\;.
\ee
Using again the Laplace method it is easy to understand that this time the integral will be dominated by the lower boundary, yielding the exponential growth rate controlled by the complexity of stable equilibria:
\be
\Xi_{\rm st}(c,\tau)=\begin{cases}
\displaystyle\Xi_{\rm tot}(c)=0&\;,\;\;c\leq c_t\;,\\
&\\
\displaystyle -\frac{\lambda_-^2}{2\,c\,\tau}+\Phi(\lambda_-,\lambda_-;c,\tau)-\int d\lambda \, n_{\mu}(\lambda)\ln|\lambda|&\;,\;\;c> c_t\;.
\end{cases}\label{xi_st_2}
\ee
Whereas the first line is obvious,  for $c>c_t$ one can show that the rate function
\be
\phi(\xi)=-\frac{\xi^2}{2\,c\,\tau}+\Phi(-\xi,-\xi;c,\tau)\;,
\ee
has a single minimum $\phi'(\xi_*)=0$, and for any $\xi>\xi_*$ holds
\be
\phi''(\xi)=\frac{1}{c\,\tau}\left[\left.\partial_x \Gamma_{c,\tau}(x)\right|_{x=-\xi}-1\right]<0\;,\label{phi_double_prime}
\ee
where the function $\partial_x \Gamma_{c,\tau}(x)$ is computed explicitly in App. \ref{AppA}.
This yields that $\phi'(\xi>\xi_*)$ is negative and the integral is indeed dominated by the lower boundary at $\xi=-\lambda_-$.

Taking the derivative with respect to $\tau$ in Eq. \eqref{xi_st_2} for $c>c_t$ and using Eq. \eqref{phi_tau}
one can show that
\begin{align}
\Xi_{\rm st}(c,\tau)&=\Xi_{\rm st}(c,1)-\int_\tau^{1}dt\,\left[\left(-\frac{\lambda_-}{c\,t}+\left.\partial_{x}\Phi\right|_{z=\bar z=\lambda_-}\right)\partial_{t}\lambda_-+\frac{\lambda_-^2}{2\,c\,t^2}+\partial_{t}\Phi(\lambda_-,\lambda_-;c,t)\right]\nn\\
&=\Xi_{\rm st}(c,1)-\frac{(1-\tau)}{2c \tau}(\lambda_--\Gamma_{c,\tau}(\lambda_-))^2\;,
\end{align}
where we have used Eqs. \eqref{Gamma_l_-} and \eqref{gamma_l_-} to verify that
\be
\partial_\tau \lambda_-=\partial_\tau\left[\Gamma_{c,\tau}(\lambda_-)\right]=\frac{\Gamma_{c,\tau}(\lambda_-)}{\tau}\;.
\ee
To simplify further this equation, we set $\tau=1$ and take a derivative with respect to $c$ in \eqref{Gamma_eq} to show that
\be
\partial_c \lambda_-=\frac{\Gamma_{c,1}(\lambda_-)}{c}\;,\;\;\tau=1\;,
\ee
which yields
\begin{align}
\partial_c\Xi_{\rm st}(c,1)&=\left(-\frac{\lambda_-}{c}+\left.\partial_{x}\Phi\right|_{z=\bar z=\lambda_-}\right)\partial_{c}\lambda_-+\frac{\lambda_-^2}{2\,c^2}+\partial_{c}\Phi(\lambda_-,\lambda_-;c,1)\nn\\
&=\left(\Gamma_{c,1}(\lambda_-)-\lambda_-\right)\frac{\Gamma_{c,1}(\lambda_-)}{c^2}+\frac{\lambda_-^2}{2\,c^2}-\frac{\Gamma_{c,1}^2(\lambda_-)}{2\,c^2}\nn\\
&=\frac{1}{2\,c^2}\left(\Gamma_{c,1}(\lambda_-)-\lambda_-\right)^2
\end{align}
Now defining the function $\nu(c)=\Gamma_{c,\tau}(\lambda_-)-\lambda_-$ which is independent of $\tau$ and satisfies
\be
\int d\lambda\,\frac{c\,n_{\mu}(\lambda)}{(\lambda+\nu(c))^2}=1\;\label{del_eq}
\ee
allows finally to represent the complexity of stable equilibria as
\be
\Xi_{\rm st}(c,\tau)=\begin{cases}
0&\;,\;\;c\leq c_t\;,\\
&\\
\displaystyle \int_{c_t}^c \frac{d\omega}{2\omega^2}\,\nu^2(\omega)-\frac{(1-\tau)}{2c \tau}\nu^2(c)&\;,\;\;c>c_t\;.
\end{cases}\label{xi_st_final}
\ee
 Such complexity obviously displays a trivialization transition by vanishing at a finite value of $\tau$ which for  a given $c>c_t$
 is given by
\be
\tau_0(c)=\frac{\nu^2(c)}{c\int_{c_t}^c \frac{d\omega}{\omega^2}\,\nu^2(\omega)+\nu^2(c)}\;,\;\,
\ee
such that $\Xi_{\rm st}(c,\tau)>0$ for $\tau>\tau_0(c)$ while $\Xi_{\rm st}(c,\tau)\leq 0$ for $\tau\leq \tau_0(c)$.
For any relaxation rate density $n_{\mu}(\lambda)$ such that the integrals $I_2(\mu)<\infty$ and
\be
I_3(\mu)=\int \frac{d\lambda}{\lambda^3}\,n_{\mu}(\lambda)<\infty\;,
\ee
are finite and well-defined, one can Taylor expand equation \eqref{del_eq} as $c\to c_t$, yielding
\be
1=\int d\lambda\,\frac{c\,n_{\mu}(\lambda)}{(\lambda+\nu(c))^2}=\frac{c}{c_t}-2c_t I_3(\mu)\nu(c)+o(c-c_t)\;.
\ee
The complexity of stable equilibria then vanishes for any $0\leq \tau\leq 1$ as
\be
\Xi_{\rm st}(c,\tau)=-\frac{(1-\tau)I_2^{5}(\mu)}{8\,\tau\,I_3(\mu)^2}(c-c_t)^2+\frac{I_2^{6}(\mu)}{24\,I_3(\mu)^2}(c-c_t)^3+o(c-c_t)^3\;,\;\;c\geq c_t\;,\label{quad_cub_van}
\ee
where we have used that $c_t=I_2(\mu)^{-1}$. Note that for any $0<\tau<1$, the complexity of stable equilibria vanishes quadratically at the transition while for $\tau=1$ it vanishes cubically. In the latter case the exponent and pre-factor match the formulae obtained in \cite{BABM21a}. Finally, setting $\Xi_{\rm st}(c,\tau_0(c))=0$, the behaviour of $\tau_0(c)$ as $c\to c_t$ reads in this regime
\be
\tau_0(c)=1-\frac{c-c_t}{3c_t}+o(c-c_t)\;.
\ee

\section{Analysis of the complexities for a power law relaxation spectrum} \label{power_law_sec}

In this section we further concentrate on analysing the complexities for the spectrum of relaxation rates described by a power law behaviour $\mu_k=\mu(k/N)^{1/\eta}$ so that the corresponding density $n_{\mu}(\lambda)$ is of the form \eqref{pow_law_spec}.
Note that the model described in \cite{SZ04} is of this type with $\eta=3/2$.

Specific features of the power law spectrum \eqref{pow_law_spec} stem from the fact that the integrals
\be
I_p(\mu)=\int_0^{\mu} \frac{d\lambda}{\lambda^p}\,n_{\mu}(\lambda)=\frac{\eta}{\eta-p}\mu^{-p}<\infty\;,\;\;\eta>p\;,
\ee
and diverge as long as $\eta\le p$. The general analysis of the previous section then immediately implies that \\
\noindent ({\bf i}) the threshold value  $c_t=I_2(\mu)^{-1}$ is positive only for $\eta>2$,\\
\noindent ({\bf ii}) the total complexity vanishes quadratically at the threshold for any $\eta>4$, see \eqref{quad_van},\\
\noindent ({\bf iii}) the complexity of stable equilibria vanishes as in \eqref{quad_cub_van} for any $\eta>3$.

Below we aim to obtaining a general expression for the total complexity and the complexity of stable equilibria for the power law spectrum and to study $\Xi_{\rm tot}(c)$ and $\Xi_{\rm st}(c,\tau)$  on approaching the transition, i.e. as $c\to c_t$ from above, in different regimes.

Before considering finite values of $\eta$, let us first briefly mention the case $\eta\to \infty$, where effectively
 $n_{\mu}(\lambda)=\delta(\lambda-\mu)$. Hence this should be equivalent to the {\it homogeneous} case $\mu_k=\mu, \forall k$ and  the results of \cite{FK16,BAFK21} should be recovered.

\subsection{Analysis of the homogeneous relaxation spectrum limit $\eta\to \infty$}

In the limit $\eta\to \infty$, computing the threshold value $c=c_t$ according to Eq. \eqref{c_t_eq} we have
\be
c_t=\left[\int \frac{d\lambda}{\lambda^2}\,n_{\mu}(\lambda)\right]^{-1}=\mu^2\;.
\ee
For any value of $c\geq \mu^2$ we can compute explicitly the integral in the right-hand side of \eqref{Upsilon} and looking for
a nonzero solution we get
\be
\int d\lambda \frac{c\,n_{\mu}(\lambda)}{\lambda^2+\Upsilon^2(c)}=\frac{c}{\mu^2+\Upsilon^2(c)}=1\;,
\ee
which yields $\Upsilon^2(c)=c-\mu^2$. The total complexity can thus be found explicitly from Eq. \eqref{xi_tot_exp_sec} as
\be
\Xi_{\rm tot}(c)=\begin{cases}
0&\;,\;\;c\leq \mu^2\;,\\
&\\
\displaystyle \frac{1}{2}\left[\frac{\mu^2}{c}-1+\ln\left(\frac{c}{\mu^2}\right)\right]&\;,\;\;c>\mu^2\;.
\end{cases}
\ee
Note that our computation here is done for a fixed $\mu$ and using $c$ as a control parameter, similarly to \cite{BABM21a} but at variance with \cite{FK16,BAFK21} where $c=1$ was fixed and $\mu$ was varying.

As in this simple case $I_p(\mu)=\int \frac{d\lambda}{\lambda^p}\,n_{\mu}(\lambda)=\mu^{-p}$ is finite for any $p>0$, the complexity vanishes quadratically as $c\to c_t=\mu^2$:
\be
\Xi_{\rm tot}(c)=\frac{1}{4\mu^4}(c-\mu^2)^2+O(c-\mu^2)^3\;,\;\;c\geq \mu^2\;.
\ee
matching exactly Eq. \eqref{quad_van}.

Considering the complexity of stable equilibria, the value of $\nu(c)$ via Eq. \eqref{del_eq} is immediate to get from
\be
\int d\lambda\,\frac{c\,n_{\mu}(\lambda)}{(\lambda+\nu(c))^2}=\frac{c}{(\mu+\nu(c))^2}=1\;,
\ee
yielding $\nu(c)=\sqrt{c}-\mu$. Inserting this expression into Eq. \eqref{xi_st_final} the complexity of stable equilibria
\be
\Xi_{\rm st}(c,\tau)=\begin{cases}
0&\;,\;\;c\leq \mu^2\;,\\
&\\
\displaystyle\Xi_{\rm tot}(c)-\frac{(1+\tau)}{2c\tau}(\mu-\sqrt{c})^2&\;,\;\;c>\mu^2\;,
\end{cases}
\ee
reproducing the behaviour found in \cite{BAFK21}. Note that the line $\tau_0(c)$ is explicit in that case:
\be
\tau_0(c)=-\frac{(\mu-\sqrt{c})^2}{\sqrt{c}\left[2(\sqrt{c}-\mu)-\sqrt{c}\ln\left(\frac{c}{\mu^2}\right)\right]}\;,\;\;c\geq \mu^2\;.\label{tau_0_triv}
\ee
Finally, the complexity of stable equilibria vanishes at $c\to c_t$ as
\be
\Xi_{\rm st}(c,\tau)=-\frac{1-\tau}{8\,\tau\,\mu^4}(c-\mu^2)^2+\frac{1}{24\mu^6}(c-\mu^2)^3+O(c-\mu^2)^4\;,\;\;c\geq \mu^2\;,
\ee
in full agreement with \eqref{quad_cub_van}. The behaviour of $\tau_0(c)$ close to the transition therefore reads
\be
\tau_0(c)=1-\frac{c-\mu^2}{3\mu^2}+O(c-\mu^2)^2\;,
\ee
as can be checked directly from Eq. \eqref{tau_0_triv}.

Note that in this homogeneous limit $\eta\to\infty$, not only the annealed total complexity and complexity of stable equilibria have been computed but also the annealed complexity for $\alpha$-stable equilibria, i.e. equilibria with $\alpha N$ stable directions \cite{BAFK21}.

\subsection{Analysis of the complexities for a  power law relaxation rate spectrum}

We will now consider the behaviour of the complexities for a power law relaxation rate spectrum with arbitrary value of $\eta>0$.

\subsubsection{Total complexity and its threshold behaviour}\hfill\\
Let us first obtain the explicit expressions for the total complexity for any value of $\mu$, $c\geq c_t$ and $\eta>0$.

We start by expressing $1/c$ from Eq. \eqref{Upsilon} with the rate density \eqref{pow_law_spec} via
\be
\frac{1}{c}=\int_0^{\mu} d\lambda \frac{n_{\mu}(\lambda)}{\lambda^2+\Upsilon^2(c)}=\frac{1}{\Upsilon^2(c)}\pFq{2}{1}{1,\frac{\eta}{2}}{\frac{\eta}{2}+1}{-\frac{\mu^2}{\Upsilon^2(c)}}\;,\label{1_c_eq}
\ee
where $\pFq{2}{1}{a,b}{c}{x}$ is the hypergeometric function. In particular, introducing the function
\be
f_{\eta}(x)=\frac{1}{x}\pFq{2}{1}{1,\frac{\eta}{2}}{\frac{\eta}{2}+1}{-\frac{1}{x}}\;,
\ee
together with its functional inverse $f_{\eta}^{-1}$, we can express conveniently $\Upsilon^2(c)$ as a scaling function of $\mu^2/c$ as
\be
\Upsilon^2(c)=\mu^2 f_{\eta}^{-1}\left(\frac{\mu^{2}}{c}\right)\;.
\ee
Taking into account that $\Upsilon^2(c_t)=0$, one then  gets the scaling function for the total complexity as
\begin{align}
\Xi_{\rm tot}(c)&=\sigma_{\eta}\left(\frac{c}{\mu^{2}}\right)\;,\;\;c\geq c_t\;,\\
\sigma_{\eta}\left(w\right)&=-\frac{1}{2}\int_{0}^{f^{-1}(w^{-1})}dx\, x\, f_{\eta}'(x)\;.\label{zeta_xi_tot}
\end{align}
{\bf Remark 4.1}. It is not at all surprising that the total complexity is a function of the ratio $c/\mu^2$ rather than $\mu$ and $c$ individually. Indeed, multiplying both matrices $D^{\mu}$ and $X^f$ by the same constant factor in Eq. \eqref{N_tot_exp} can not change the number of total equilibria. However, it does depend on the ratio between typical scales for eigenvalues of matrices $D^{\mu}$ and $X^{f}$, the former being controlled by $\mu$ while the latter by $\sqrt{c}$.

 The large argument  behaviour of the scaling function $\sigma_{\eta}(w)$ is quite straightforward to obtain:
\be
\sigma_{\eta}(w)=\frac{1}{2}\ln w-\frac{1}{2}+\frac{1}{\eta}+o(1)\;.\label{large_w_sig}
\ee
On the other hand, the behaviour of this function at the threshold as $c\to c_t=\lim_{x\to 0}\mu^{2}/f_\eta(x)$, will depend nontrivially on the value of the exponent $\eta$. In order to obtain the leading order behaviour of the scaling function $\sigma_{\eta}(w)$, we need the function $f_{\eta}(x)$ for small values of $x$. The corresponding expansion reads
\be
f_{\eta}(x)=\begin{cases}
\displaystyle \frac{\eta}{\eta-2}+\frac{\eta}{\eta-4}x+\frac{\eta \pi}{2\sin\left(\frac{\eta \pi}{2}\right)}x^{\frac{\eta-2}{2}}+O(x^2)&\;,\;\;\eta\neq 2,4\;,\\
&\\
\displaystyle -\ln x+x+O(x^2)&\;,\;\;\eta= 2\;,\\
&\\
\displaystyle 2+2x\ln x+O(x^2)&\;,\;\;\eta= 4\;,
\end{cases}
\ee
where we remind that $c_t=\lim_{x\to 0} \mu^{2}/f_{\eta}(x)$. Clearly, in the first line there is a competition between the terms $O(x)$ and $O(x^{\frac{\eta-2}{2}})$, with the leading order behaviour for $\eta>4$ being controlled by the former and for $\eta<4$ by the latter. Using it we can compute explicitly the leading order behaviour of $f_{\eta}^{-1}(\kappa)$ as $\kappa\to \lim_{x\to 0}f_{\eta}(x)$, and  the corresponding scaling function $\sigma_{\eta}(w)$.  The ensuing leading order behaviour of the total complexity $\Xi_{\rm tot}(c)$ as $c\to c_t$ is then given exactly by the equation \eqref{xi_tot_res}. In particular, it is evident that as long as $0<\eta\leq 4$ the threshold behaviour depends explicitly on $\eta$ and is quite different from the quadratic behaviour in Eq. \eqref{quad_van}.

Finally, let us define for finite $N$ the two quantities
\be
\Xi_{N,{\rm SA}}(c)=\frac{1}{N}\moy{\ln|\det(D^{\mu}-X)|}_{\rm rEGE}\;,\;\;\Xi_{N}(c)=\frac{1}{N}\ln\moy{|\det(D^{\mu}-X)|}_{\rm rEGE}\;,
\ee
Assuming the self-averaging hypothesis to hold for $D^{\mu}$ with power law spectrum we should expect that both $\Xi_{N,{\rm SA}}(c)$ and $\Xi_{N}(c)$ converge as $N\to\infty$ to the same scaling function $\sigma_{\eta}(c/\mu^2)$. In Fig. \ref{tot_comp}, we show a comparison between these two quantities for $N=10^3$ and choosing $\mu=1$, $\tau=0$
and $\eta=6,3,1$ respectively. The agreement is excellent and shows a transition for finite $c_t=2/3,1/3$ for $\eta=6,3$ respectively while the transition occurs for $c_t=0$ for $\eta=1$.

\begin{figure}
\centering
\includegraphics[width=0.48\textwidth]{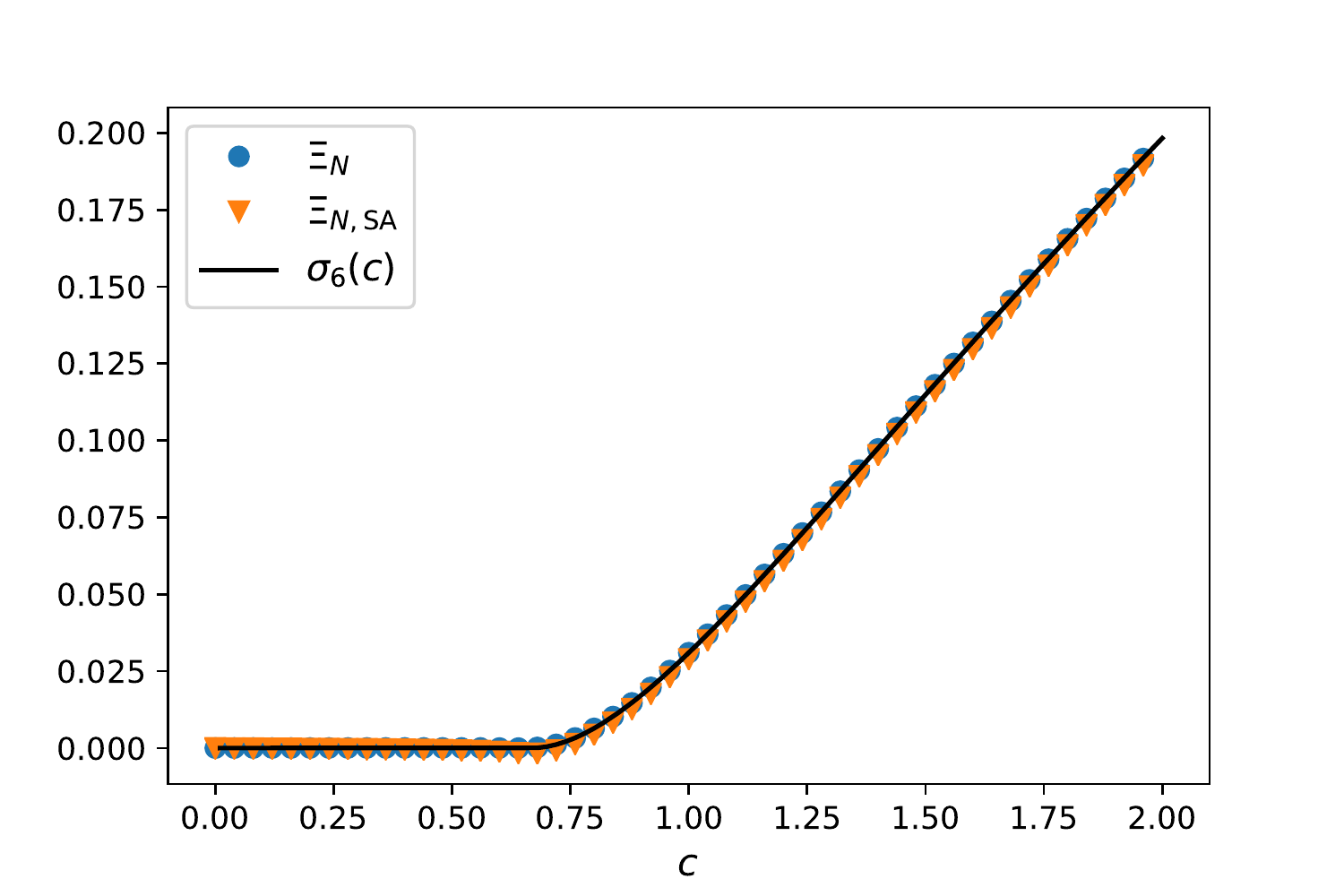}
\includegraphics[width=0.48\textwidth]{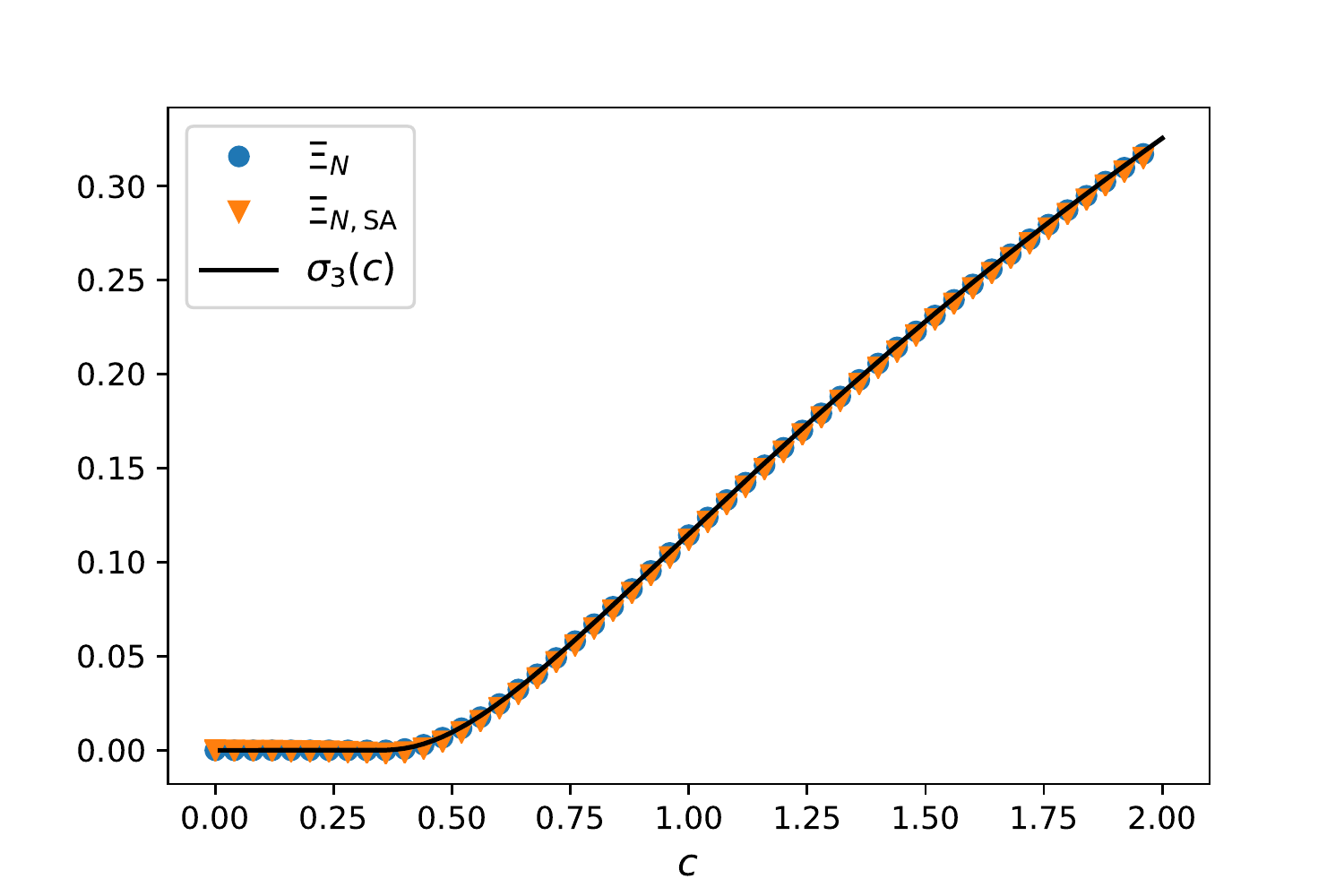}
\includegraphics[width=0.48\textwidth]{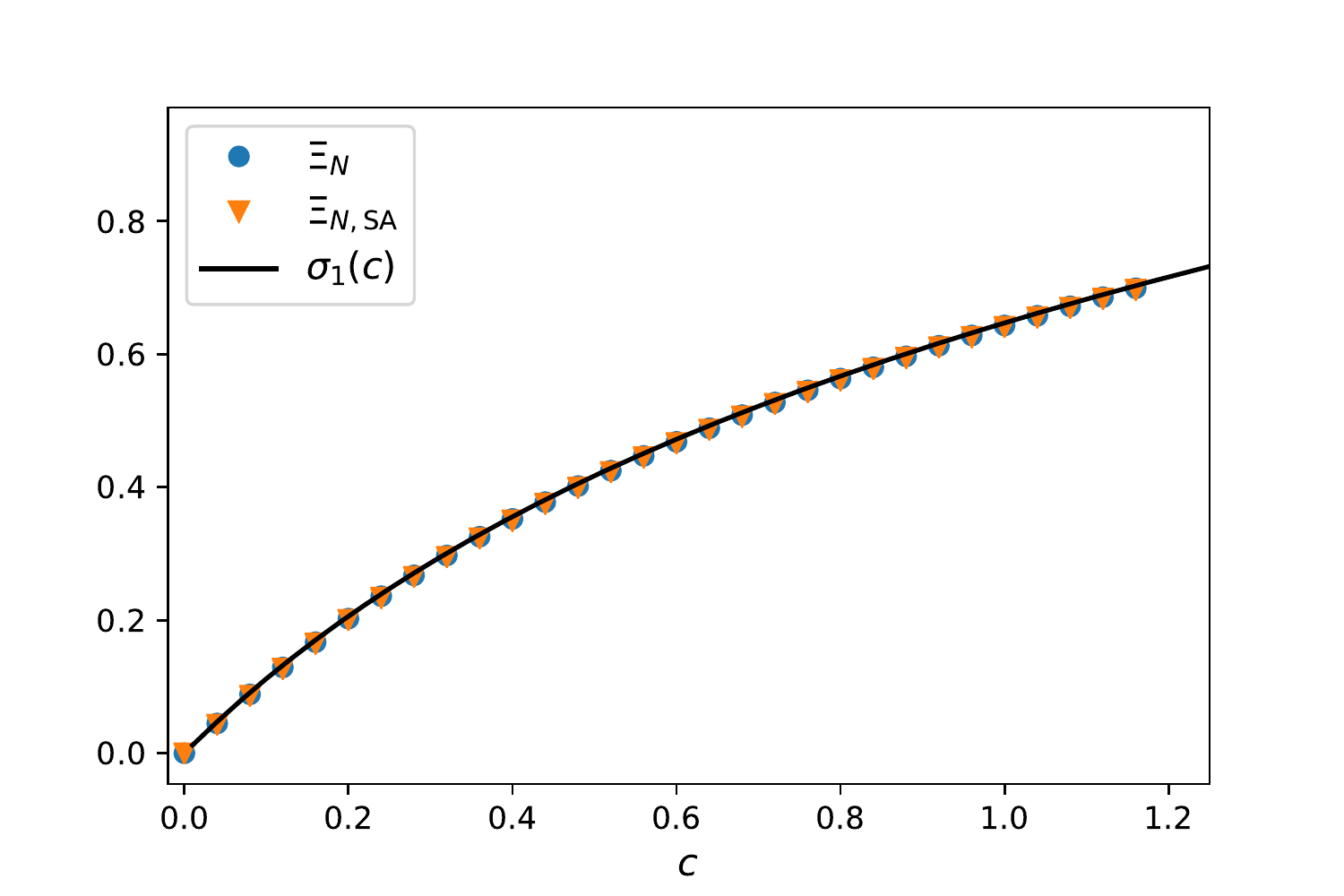}
\caption{Comparison between the finite $N$ quantities $\Xi_{N,{\rm SA}}(c)=\frac{1}{N}\moy{\ln|\det(D^{\mu}-X)|}_{\rm rEGE}$ (orange triangles) and $\Xi_{N}(c)=\frac{1}{N}\ln\moy{|\det(D^{\mu}-X)|}_{\rm rEGE}$ (blue dots) for $\mu=1$, $\tau=0$ and $N=10^3$ and the scaling function $\sigma_{\eta}(c)$ (black line). The upper-left case corresponds to a power-law spectrum with exponent $\eta=6$, the upper right to $\eta=3$ and lower centre to $\eta=1$. In each case, the three quantities match perfectly. The transition from the trivial to the complex phase occurs for $c=c_t=\max((\eta-2)/\eta,0)$. At the transition, the complexity behaves quadratically for $\eta=6$, cubically for $\eta=3$ and linearly for $\eta=1$.}\label{tot_comp}
\end{figure}

\subsection{Complexity of stable equilibria and its threshold behaviour}

We consider now the complexity of stable equilibria for the power law spectrum in Eq. \eqref{pow_law_spec} for any value of $\mu$, $c\geq c_t$ and $\eta>0$. Let us first show how to obtain an exact expression for this complexity.

 We start by considering the expression of $1/c$ obtained by inserting the spectral density \eqref{pow_law_spec} in Eq. \eqref{del_eq}. It reads
\be
\frac{1}{c}=\int_0^{\mu} d\lambda \frac{n_{\mu}(\lambda)}{(\lambda+\nu(c))^2}=\frac{1}{\nu^2(c)}\pFq{2}{1}{2,\eta}{\eta+1}{-\frac{\mu}{\nu(c)}}\;.\label{1_c_eq_del}
\ee
Introducing the function
\be
g_{\eta}(x)=\frac{1}{x^2}\pFq{2}{1}{2,\eta}{\eta+1}{-\frac{1}{x}}\;,
\ee
together with its functional inverse  $g_{\eta}^{-1}$, one can express $\nu(c)$ explicitly as a function of $c$ as
\be
\nu(c)=\mu\, g_\eta^{-1}\left(\frac{\mu^{2}}{c}\right)\;.
\ee
Using this, the complexity of stable equilibria can be expressed as the following scaling function
\begin{align}
\Xi_{\rm st}(c,\tau)&=\chi_{\eta}\left(\frac{c}{\mu^{2}};\tau\right)\;,\\
\chi_{\eta}(w;\tau)&=-\frac{1}{2}\int_0^{g_\eta^{-1}\left(w^{-1}\right)}du\,u^2\,g_\eta'(u)\;-\frac{(1-\tau)}{2\,w\,\tau}g_\eta^{-1}\left(\frac{1}{w}\right)^2\;.
\end{align}
The asymptotic $w\to \infty$ behaviour of the scaling function $\chi_{\eta}(w;\tau)$ is quite simple to obtain and reads
\be
\chi_{\eta}(w;\tau)=\frac{1}{2}\ln w-\frac{3}{2}+\frac{1}{\eta}+o(1)\;.
\ee
Note in particular that this expression is independent of $\tau$ and $\sigma_{\eta}(w)-\chi_{\eta}(w;\tau)=\Xi_{\rm tot}(c)-\Xi_{\rm st}(c,\tau)=1+o(1)$ is independent of both $c$ and $\tau$ at the leading order.

 The behaviour of the complexity of stable equilibria close to the threshold is controlled by the behaviour of the function $g_{\eta}(x)$ at small arguments. As $x\to 0$ one gets the expansion:
\be
g_{\eta}(x)=\begin{cases}
\displaystyle \frac{\eta}{\eta-2}-\frac{2\eta}{\eta-3}x-\frac{\eta (\eta-1) \pi}{\sin\left(\eta \pi\right)}x^{\eta-2}+O(x^2)&\;,\;\;\eta\neq 2,3\;,\\
&\\
\displaystyle -2(\ln x+1)+4x+O(x^2)&\;,\;\;\eta= 2\;,\\
&\\
\displaystyle 3+3x(2\ln x+x)+O(x^2)&\;,\;\;\eta= 3\;,
\end{cases}
\ee
where we remind that $c_t=\lim_{x\to 0} \mu^{2}/f_{\eta}(x)=\lim_{x\to 0} \mu^{2}/g_{\eta}(x)$. As for the total complexity, the first line of this equation indicates a clear change in the leading behaviour where the term $O(x^{\eta-2})$ dominates for $\eta<3$ while the term $O(x)$ dominates for $\eta>3$. Using the behaviour of the function $g_{\eta}(x)$, we can compute the leading order behaviour of $g_{\eta}^{-1}(\kappa)$ as $\kappa\to \lim_{x\to 0}g_{\eta}(x)$ and the expression of the scaling function $\chi_{\eta}(w;\tau)$. This yields the expression of the complexity of stable equilibria as $c\to c_t$ provided in the equation \eqref{xi_stab_res}.
In particular, the complexity of stable equilibria vanishes with an exponent that depends explicitly on the value of $\eta$ in the regime $\eta\leq 3$. As a by-product of this computation we also obtain the behaviour of $\tau_0(c)$ as $c\to c_t$. For any $\eta> 2$, we obtain that $\tau_0(c)\to 1$ at the leading order as $c\to c_t$, while $\tau_0(c)\to \eta/2$ for any $\eta<2$.

\section{Conclusion}\label{conclu}

In this article, we have considered the (annealed) complexities of total equilibria and  of stable equilibria for a random  system \eqref{evol_xi} of $N$ coupled autonomous ordinary differential equations in the limit $N\gg1$.
 Assuming that the couplings between different degrees of freedom are provided by a Gaussian translationally invariant random vector field and using the Kac-Rice formalism this problem is conveniently re-formulated in terms of the averaged modulus of the determinant of the random Jacobian associated to this field.

Further assuming the self-averaging property of this large determinant \eqref{detselfaver} we have obtained exact formulae for both these annealed complexities extending the known rigorous results \cite{FLD20_2,BABM21a} for gradient random field to fields with both gradient and solenoidal components. We have shown that the total complexity is independent of the fraction $\tau$ of gradient components while the complexity of stable equilibria undergoes an additional transition: it is negative for $\tau\leq \tau_0(c)$  and positive conversely.

Finally, we have analysed the behaviour of the complexities close to the tivialisation transition for a limiting density of the rates behaving as $n_{\mu}(\lambda)\sim \lambda^{\eta-1}$ as $\lambda\to 0$. The quadratic vanishing $\Xi_{\rm tot}(c)\propto (c-c_t)^2$ of the total complexity observed for a gaped density extends for any $\eta>4$ but for $\eta\leq 4$, it vanishes with an exponent depending explicitly on $\eta$.

As mentioned in the introduction, disordered elastic manifolds constitute another natural physical application of the general framework exposed in this article. In particular, our analytical results for power law relaxation spectrum give interesting insights on disordered elastic manifold of internal dimension $d\geq 4$. This will be the object of a future publication.

The analysis of Gaussian random field beyond the "zero-range model" with correlated covariance tensor $C_{ijkl}\neq C_{ijkl}(\tau)$, like in Eq.\eqref{cov_3}, is a much more difficult task that we hope to address in a future publication.
Let us nevertheless stress that our Kac-Rice based formalism allows to put evaluation of the number of solutions on the firm numerical ground. In particular, in the motivating case of the Spivak-Zyuzin model this allowed us to demonstrate that a simple-minded 
growth rate estimate proposed in \cite{SZ04} can not correctly account for the actual behaviour of the model.
Building a more complete understanding of the counting problem for such a model remains therefore an interesting challenge.

\newpage

\appendix

\section{Derivation of the mean eigenvalue density for a real diagonal perturbation of the real Elliptic Gaussian Ensemble} \label{AppA}

Let us define the matrix
\be
J=D^{\mu}+X\;,
\ee
where $D_{ij}^{\mu}=\mu_i \delta_{ij}$ is a prescribed real diagonal matrix for which we know its limiting spectral density
\be
n_{\mu}(\lambda)=\lim_{N\to \infty}\frac{1}{N}\sum_{k=1}^{N}\delta(\mu_{k}-\lambda)\;,
\ee
 and $X$ is a matrix drawn from the real Gaussian Elliptic Ensemble with
\be
\moy{X_{ij}}=0\;,\;\;\moy{X_{ij}X_{kl}}=\frac{c}{N}(\delta_{ik}\delta_{jl}+\tau\,\delta_{il}\delta_{jk})\;.
\ee
Our aim is to obtain, in the limit $N\to \infty$, the closed-form expressions for the mean density of its complex eigenvalues
\be
\rho(z,\bar z)=\lim_{N\to \infty}\frac{1}{N}\sum_{k=1}^{N}\moy{\delta(z_{k}-z)\delta(\bar z_{k}-\bar z)}=\frac{2}{\pi}\partial_{z \bar z}^2 \Phi(z,\bar z;c,\tau)
\ee
with the associated electrostatic potential
\begin{align}\label{potential}
\Phi(z,\bar z;c,\tau)&=\lim_{N\to \infty}\frac{1}{2N}\moy{\Tr \ln\left[(\bar z\mathbb{I}- H^{\dagger})(z\mathbb{I}-H)\right]}\\
&=\frac{1}{2}\int d^2 w \, \rho(w,\bar w)\ln|(z-w)(\bar z-\bar w)|\;.\nn
\end{align}
as the latter plays an important role in the computation of the total complexity in Eq. \eqref{xi_tot_2} and of the complexity of stable equilibria in Eq. \eqref{xi_st_2}.

The main result of this appendix is detailed in the Proposition formulated in Sec. (\ref{RMTsec}).
Below we  employ two alternative ways of verifying the Proposition, the first  relying on the integration over anticommuting variables and the second follows closely a similar derivation of \cite{K96} where the special case $\tau=0$ has been treated.

\subsection{Derivation via strong self-averaging and Grassmann integration}

The starting point in this approach is again the assumption that the strong self-averaging property of the logarithm holds for
 the (regularized) electrostatic potential \eqref{potential}, namely that the operations of taking the ensemble average
and taking the logarithm commute in the limit $N\to \infty$, cf. \eqref{detselfaver}:
\begin{align}\label{potential_selfave}
\Phi(z,\bar z;c,\tau)&=\lim_{N\to \infty}\frac{1}{2N}\moy{\Tr \ln\left[(\bar z\mathbb{I}- J^{\dagger})(z\mathbb{I}-J)\right]}\\
&=\lim_{N\to \infty}\frac{1}{2N} \ln\moy{\det\left[(\bar z\mathbb{I}- J^{\dagger})(z\mathbb{I}-J)\right]}\;.\label{potential_selfaveB}
\end{align}
A possibility to use a similar commutativity for deriving the expressions for the mean density of eigenvalues was first noticed in \cite{Ber73} for Hermitian random matrices. Note that in a few classical cases of non-Hermitian matrices the above identity can be rigorously verified, e.g. for the complex Ginibre matrices,  see eq. (2.19) of \cite{FK07a}. Exploiting \eqref{potential_selfave} provides a powerful basis for an efficient calculation of the eigenvalue densities in nontrivial cases, like e.g. in the ''single-ring'' class \cite{FZ97,GKZ11} of non-Hermitian ensembles, see \cite{FK07b}.
We therefore simply conjecture its validity for the perturbed real elliptic Ginibre case and aim at evaluating
\begin{equation}\label{DNdef}
\moy{\det\left[(\bar z\mathbb{I}- J^{\dagger})(z\mathbb{I}-J)\right]}=\moy{\det\left(
\begin{array}{cc}\mathbb{O} & i\left(z-J\right)\\i\left(\bar z-J^T\right)&  \mathbb{O}\end{array}\right)}
\end{equation}
where we used that we are dealing with the real-valued matrices.

The block-offdiagonal determinant in \eqref{DNdef} can be conveniently represented by a Gaussian integral over
anticommuting Grassmann variables. Namely,
let $\Psi_1,\,\Psi_2,\,\Phi_1,\,\Phi_2$ be four column vectors with $N$ anticommuting components each. Using the standard rules of
Berezin integration we write
\begin{equation*}\label{detGrassmanGauss}
\det{\begin{pmatrix}
  \mathbb{O}  & i(z\,\mathbb{I}-J) \\
  i(\bar z\,\mathbb{I}-J^T) &  \mathbb{O}
\end{pmatrix}}
 = \int d\Psi_1\,d\Psi_2\,d\Phi_1\,d\Phi_2 \,\,  e^{-i(\Psi^T_1,\Phi^T_1)\begin{pmatrix}
   \mathbb{O}   & z\,\mathbb{I} -J \\
  \bar z \,\mathbb{I}-J^T &  \mathbb{O}
\end{pmatrix}\begin{pmatrix}\Psi_2 \\ \Phi_2\end{pmatrix} }\,.
\end{equation*}
\begin{equation*}
 = \int d\Psi_1\,d\Psi_2\,d\Phi_1\,d\Phi_2 \,\,  e^{-i\Psi^T_1( z\,\mathbb{I} -D^{\mu})\Phi_2-i\Phi^T_1(\bar z\,\mathbb{I} -D^{\mu})\Psi_2  }\,.
\end{equation*}
\begin{equation}\label{detGrassmanGauss}
 \times  e^{i\Psi^T_1 X \Phi_2+i\Phi^T_1 X^T \Psi_2 }\,.
\end{equation}
and perform the averaging over elliptic ensemble matrices $V$ using the identity
\begin{equation}\label{Gauident}
\left\langle
e^{-i\Tr(XA+X^TB)}
\right\rangle_{rEGE}=e^{-\frac{c}{2N}\Tr(A^TA+B^TB+2AB)-\frac{c\tau}{2N}\Tr(A^2+B^2+2AB^T)}\,,
\end{equation}
where in our case $A=\Phi_2\otimes \Psi_1^T$ and $B=\Psi_2\otimes \Phi_1^T$. This implies
\begin{equation}\label{aveexp}
 \moy{ e^{i\Psi^T_1 X \Phi_2+i\Phi^T_1 X^T \Psi_2 }}_{rEGE}=e^{\frac{c}{2N}(\Psi_1^T\Psi_2)(\Phi_1^T\Phi_2)
 +\frac{c\tau}{2N}\left[(\Psi_1^T\Phi_2)^2+(\Phi_1^T\Psi_2)-2(\Psi_1^T\Phi_1)(\Psi_2^T\Phi_2)\right]}\,.
\end{equation}
Using the set of Hubbard-Stratonovich decouplings:
\begin{align}\label{HSgrasreal}
e^{ \frac{c\tau}{2N}(\Psi_1^T\Phi_2)^2}&=\sqrt{\frac{N}{2c\pi}}\int_{\mathbb{R}}e^{-\frac{Nu_1^2}{2c}+u_1\sqrt{\tau}(\Psi_1^T\Phi_2)}du_1\;,\\
\quad e^{ \frac{c\tau}{2N}(\Phi_1^T\Psi_2)^2}&=\sqrt{\frac{N}{2c\pi}}\int_{\mathbb{R}}e^{-\frac{Nu_2^2}{2c}+u_2\sqrt{\tau}(\Phi_1^T\Psi_2)}du_2\;,
\end{align}
as well as
\begin{equation}\label{HSgrascompA}
e^{ \frac{c\tau}{2N}(\Psi_1^T\Phi_2)^2}=\frac{N}{2\pi\, c}\int_{\mathbb{C}}e^{-\frac{N}{c}\bar b\,b-i\sqrt{\tau}
\left[b(\Psi_1^T\Phi_1)+\bar b(\Psi_1^T\Phi_1)\right]}db\,d\bar b,
\end{equation}
and finally
\begin{equation}\label{HSsimple}
e^{\frac{c}{2N}\left(\Phi_1^T\Phi_2\right)\left(\Psi_1^T\Psi_2\right)}=\frac{1}{2\pi}\int_{\mathbb{C}} e^{-\frac{N}{c}\bar q\,q-q\left(\Psi_1^T\Psi_2\right)-\bar q\left(\Phi_1^T\Phi_2\right)} dq\, d\bar q \,,
\end{equation}
after simple rearranging and change of integration orders one can integrate out Grassmann variables explicitly by using the Pfaffian identity
\begin{equation}\label{gauintpfaff}
\int e^{-\frac{1}{2}\zeta^TC\zeta}{\cal D}\zeta=Pf(C)
\end{equation}
where $C$ is any antisymmetric matrix and $\zeta$ is a vector with anticommuting Grassmannian components. In our case $\zeta^T:=(\Psi_1,\Phi_1,\Psi_2,\Phi_2)^T$ and
\begin{equation}
C=\left(\begin{array}{cccc} \mathbb{O} & ib\sqrt{\tau}\,\mathbb{I} &q\,\mathbb{I}& i(z\mathbb{I}-D^{\mu}+iu_1\,\sqrt{\tau}\,\mathbb{I}) \\
-ib\sqrt{\tau}\,\mathbb{I} & \mathbb{O} & i(\bar z\mathbb{I}-D^{\mu}+iu_2\,\sqrt{\tau}\,\mathbb{I})& \bar q\,\mathbb{I}\\
q\,\mathbb{I}& -i(\bar z\mathbb{I}-D^{\mu}+iu_2\,\sqrt{\tau}\,\mathbb{I}) & \mathbb{O} & i\bar b\sqrt{\tau}\,\mathbb{I}\\
-i(z\mathbb{I}-D^{\mu}+iu_1\,\sqrt{\tau}\,\mathbb{I})&-\bar q\,\mathbb{I}&-i\bar b\sqrt{\tau}\,\mathbb{I}&\mathbb{O}
\end{array}\right)
\end{equation}
Combining all the contributions we therefore have for the electrostatic potential via \eqref{potential_selfave}-\eqref{potential_selfaveB}
\begin{align}\label{potential_selfaveC}
\Phi(z,\bar z;c,\tau)&=\lim_{N\to \infty}\frac{1}{2N} \ln {\cal K}_N, \quad
{\cal K}_N=\int_{\mathbb{C}^2}\frac{dqd\bar q}{\pi}\frac{dbd\bar b}{\pi}\int_{\mathbb{R}^2}\frac{du_1du_2}{2\pi}e^{-N{\cal L}}
\end{align}
where
\be \label{action1}
{\cal L}=\frac{q\bar q+b\bar b}{c}+\frac{u_1^2+u_2^2}{2c}-\frac{1}{N}\sum_{k=1}^N\ln{W_i}
\ee
with
\be \label{action2}
W_k=q\bar q+\tau \, b\bar b+\left(z-\mu_k+iu_1\,\sqrt{\tau}\right)\left(\bar z-\mu_k+iu_2\,\sqrt{\tau}\right)\;.
\ee
Passing to the polar coordinates: $q=r_1e^{i\theta_1}, \, b=r_2e^{i\theta_2}$ with $r_{1,2}\ge 0$ and $0\le \theta_{1,2}<2\pi$ one may
 evaluate the integrals in \eqref{potential_selfaveC} for $N\gg 1$ by the Laplace method. In doing so we need to assume that
 for $0\le \tau<1$. One then finds that
the relevant saddle-point is either a trivial one $r_1=r_2=0$ (which we do not consider below), or is given by
  $r_1:=r>0$ and $r_2=0$ where $r$ satisfies
\be\label{sp1}
1=c\int_{\mathbb{R}}\frac{n_{\mu}(\lambda)\,d\lambda}{r^2+\left(z-\lambda+iu_1\,\sqrt{\tau}\right)\left(\bar z-\lambda+iu_2\,\sqrt{\tau}\right)}
\ee
whereas $u_{1,2}$ satisfy the system of two equations
\be\label{sp2}
u_1=i\sqrt{\tau}\left[\bar z+iu_2\sqrt{\tau}-c\int_{\mathbb{R}}\frac{n_{\mu}(\lambda)\,\lambda\,d\lambda}{r^2+\left(z-\lambda+iu_1\,\sqrt{\tau}\right)\left(\bar z-\lambda+iu_2\,\sqrt{\tau}\right)}\right]
\ee
and
 \be\label{sp3}
u_2=i\sqrt{\tau}\left[z+iu_1\sqrt{\tau}-c\int_{\mathbb{R}}\frac{n_{\mu}(\lambda)\,\lambda
\,d\lambda}{r^2+\left(z-\lambda+iu_1\,\sqrt{\tau}\right)\left(\bar z-\lambda+iu_2\,\sqrt{\tau}\right)}\right]\,.
\ee
It is natural to replace $u_{1,2}$ with the combinations
\be\label{sp4}
p_1=z+iu_1\sqrt{\tau}, \quad p_2=\bar z+iu_2\sqrt{\tau}
\ee
so that the relations \eqref{sp1}-\eqref{sp3} take the form
\be\label{sp1a}
1=c\int_{\mathbb{R}}\frac{n_{\mu}(\lambda)\,d\lambda}{r^2+\left(p_1-\lambda\right)\left(p_2-\lambda\right)}
\ee
\be\label{sp2a}
z-p_1=\tau\left[p_2-c\int_{\mathbb{R}}\frac{n_{\mu}(\lambda)\,\lambda\,d\lambda}{r^2+\left(p_1-\lambda\right)\left(p_2-\lambda\right)}\right]
\ee
and
 \be\label{sp3a}
\bar z-p_2=
\tau\left[p_1-c\int_{\mathbb{R}}
\frac{n_{\mu}(\lambda)\,\lambda\,d\lambda}{r^2+\left(p_1-\lambda\right)\left(p_2-\lambda\right)}\right]
\ee
which shows that $p_1$ and $p_2$ are linearly related:
\be\label{sp4a}
p_2=p_1-\frac{z-\bar z}{1-\tau}
\ee
and hence introducing $q=p_1-\frac{z-\bar z}{2(1-\tau)}$ one can write
\[
\left(p_1-\lambda\right)\left(p_2-\lambda\right)=\left(q-\lambda\right)^2+\frac{(z-\bar z)^2/4}{(1-\tau)^2}
\]
and finally introducing $z=x+iy$
after straightforward algebraic manipulations we get a closed system for $r$ and $q$:
\be\label{spb}
1=c\int_{\mathbb{R}}\frac{n_{\mu}(\lambda)\,d\lambda}{r^2+\frac{y^2}{(1-\tau)^2}+(q-\lambda)^2}, \quad
\frac{q(1+\tau)-x}{\tau}=c\int_{\mathbb{R}}\frac{n_{\mu}(\lambda)\,\lambda\,d\lambda}{r^2+\frac{y^2}{(1-\tau)^2}+(q-\lambda)^2}\,.
\ee
Note that as is easy to see the solution $q$ of this system is $y-$independent: $q=q(x)$.

The corresponding electrostatic potential is given from \eqref{potential_selfaveC} by
\be\label{Phi_el1}
\Phi(z,\bar z;c,\tau)=-\frac{r^2}{c}+\frac{1}{2\tau c}\left[(p_1-z)^2+(p_2-\bar z)^2\right]+\int_{\mathbb{R}}d\lambda\, n_{\mu}(\lambda)
\ln{\left[r^2+\left(p_1-\lambda\right)\left(p_2-\lambda\right)\right]}
\ee
and using the stationarity equations we see that
\[
\frac{\partial}{\partial z}\Phi(z,\bar z;c,\tau)=\frac{p_1-z}{c\tau},
\]
hence
\[
\rho(z,\bar z)=\frac{1}{\pi}\frac{\partial^2}{\partial z\partial \bar z}
\Phi(z,\bar z;c,\tau)=-\frac{1}{c\tau}\frac{\partial p_1}{\partial \bar z}
\]
which finally expresses the mean density of eigenvalues in the complex plane $z=x+iy$ via the solution of \eqref{spb}
as
\be\label{den_el1}
\rho(x,y)=\frac{1}{2\pi c\tau}\left[\frac{1}{1-\tau}-\frac{\partial q}{\partial x}\right]
\ee
We will continue the analysis of such a density leading to the content of the {\bf Proposition}
after giving an alternative derivation of \eqref{spb}.

\subsection{Alternative derivation following the method of \cite{K96}}

Let us first define a resolvent matrix $G(z,\bar z;\kappa)$ and its mean normalized trace $R(z,\bar z;\kappa)$ via
\be
G(z,\bar z;\kappa)=((\bar z\mathbb{I}- J^{\dagger})(z\mathbb{I}-J)+\kappa^2 \mathbb{I})^{-1}\;,\;\;R(z,\bar z;\kappa)=\frac{1}{N}\moy{\Tr G}\;.
\ee
Note that by its  definition this matrix is ({\bf i}) self-adjoint: $G^{\dagger}=G$ and ({\bf ii}) related to  the electrostatic potential  as
\begin{align}\label{dphi_z}
\partial_{z}\Phi(z,\bar z;c,\tau)&=\frac{1}{2}\lim_{\kappa\to 0} Q(z,\bar z;\kappa)\;,\quad
\partial_{\bar z}\Phi(z,\bar z;c,\tau)=\frac{1}{2}\lim_{\kappa\to 0}\bar Q(z,\bar z;\kappa)\,,\\
\mbox{where} \quad & Q(z,\bar z;\kappa)=\frac{1}{N}\moy{\Tr\left[G(\bar z\mathbb{I}-J^{\dagger})\right]}\;.
\end{align}
The mean density can therefore be obtained as
\be
\rho(z,\bar z)=\frac{1}{\pi}\lim_{\kappa\to 0} \partial_{\bar z}Q(z,\bar z;\kappa)=\frac{1}{\pi}\lim_{\kappa\to 0} \partial_{z}\bar Q(z,\bar z;\kappa)\;.
\ee

Our starting point is the following identity, see Eq. (4) in \cite{K96} that we reproduce below:
\be
\kappa^2 \moy{G}=\mathbb{I}-\overline{(z\mathbb{I}-D^{\mu})}\moy{(z\mathbb{I}-J)G}+\moy{X^{\dagger}(z\mathbb{I}-J)G}\;.\label{G_eq}
\ee
Note that in our case the matrices $X$ and $D^{\mu}$ are real. We first want to evaluate
\be
\moy{(z\mathbb{I}-J)G}=(z\mathbb{I}-D^{\mu})\moy{G}-\moy{X G}\;.
\ee
Let us first compute
\begin{align}
\moy{[XG]_{ij}}=&\sum_{k}\moy{X_{ik}G_{kj}}=\sum_{k,l,m}\moy{X_{ik}X_{lm}}\moy{\partial_{X_{lm}}G_{kj}}\\
=&\frac{c}{N}\left(\moy{\partial_{X_{ik}}G_{kj}}+\tau\,\moy{\partial_{X_{ki}}G_{kj}}\right)\;.\nn
\end{align}
Using the identity \cite{K96}
\be
\partial_{X_{kl}}G_{ij}=[G(\bar z \mathbb{I}-J^{\dagger})]_{ik}G_{lj}+G_{il}[(z \mathbb{I}-J)G]_{kj}\;,\label{nice_id}
\ee
one can show that
\begin{align}
\nn\\
\moy{[XG]_{ij}}=&\frac{c}{N}\sum_k\left[\moy{[G(\bar z\mathbb{I}-J^{\dagger})]_{ki}G_{kj}}+\moy{G_{kk}[(z\mathbb{I}-J)G]_{ij}}\right]\\
&+\frac{c\,\tau}{N}\,\sum_k\left[\moy{[G(\bar z\mathbb{I}-J^{\dagger})]_{kk}G_{ij}}+\moy{G_{ki}[(z\mathbb{I}-J)G]_{kj}}\right]\nn\\
= &\frac{c}{N}\left(\moy{\Tr[G][(z\mathbb{I}-J)G]_{ij} }+\tau\moy{G_{ij}\Tr\left[G(\bar z\mathbb{I}-J^{\dagger})\right]}\right)+O(N^{-1})\;.\nn
\end{align}
In this equation, the term $G(\bar z\mathbb{I}-J^{\dagger})=\left[(z\mathbb{I}-J)G\right]^{\dagger}$ appears. It only appears through its trace which satisfies
\be
\Tr\left[G(\bar z\mathbb{I}-J^{\dagger})\right]=\overline{\Tr[(z\mathbb{I}-J)G]}\;.
\ee
We will now introduce the main approximation needed to obtain the average density in closed form, namely that the traces are self-averaging quantities such that for large $N$
\begin{align}
&\frac{1}{N}\Tr[G]=R(z,\bar z;\kappa)+O(N^{-1})\;,\\
&\frac{1}{N}\Tr\left[G(\bar z\mathbb{I}-J^{\dagger})\right]=Q(z,\bar z;\kappa)+O(N^{-1})\;.\nn
\end{align}
Using these approximations, we obtain that
\be
\moy{(z\mathbb{I}-J)G}=(z\mathbb{I}-D^{\mu})\moy{G}-c\,R \moy{(z\mathbb{I}-J)G}-c\,\tau\,Q\moy{G}+O(N^{-1})\;.
\ee
Taking the trace on both side of the equations, we obtain an equation for $Q(z,\bar z;\kappa)$ as a function of its complex conjugate $\bar Q(z,\bar z;\kappa)$, of $R(z,\bar z;\kappa)$ and the function
\be
Q_0(z,\bar z;\kappa)=\frac{1}{N}\moy{\Tr\left[G(\bar z\mathbb{I}-D^{\mu})\right]}\;.
\ee
where we used that $D^{\mu}$ is self-adjoint.
This yields in particular
\begin{align}
\Re[Q(z,\bar z;\kappa)]=&\frac{\Re[Q_0(z,\bar z;\kappa)]}{1+c (1+\tau)R(z,\bar z;\kappa)}\;,\label{re_Q}\\
\Im[Q(z,\bar z;\kappa)]=&\frac{\Im[Q_0(z,\bar z;\kappa)]}{1+c(1-\tau)R(z,\bar z;\kappa)}\;.\label{im_Q}
\end{align}
Finally, let us compute the last term that appears in \eqref{G_eq}
\begin{align}
\moy{[X^{\dagger}(z\mathbb{I}-J)G]_{ij}}=&\sum_{k}\moy{X_{ki}[(z\mathbb{I}-J)G]_{kj}}\\
=&c\sum_{k,l}\left(\moy{\partial_{X_{ki}}[(z\mathbb{I}-J)_{kl}G_{lj}]}+\tau\,\moy{\partial_{X_{ik}}[(z\mathbb{I}-J)_{kl}G_{lj}]}\right)\nn\\
=&-c\left(1+\frac{\tau}{N}\right)\moy{G_{ij}}\nn\\
&+\frac{c}{N}\sum_{k,l}\moy{(z\mathbb{I}-J)_{kl}\left([G(\bar z\mathbb{I}-J^{\dagger})]_{lk}G_{ij}+G_{li}[(z\mathbb{I}-J)G]_{kj}\right)}\nn\\
&+\frac{c\,\tau}{N}\sum_{k,l}\moy{(z\mathbb{I}-J)_{kl}\left([G(\bar z\mathbb{I}-J^{\dagger})]_{li}G_{kj}+G_{lk}[(z\mathbb{I}-J)G]_{ij}\right)}\nn\\
=&-c\moy{G_{ij}}+\frac{c}{N}\moy{\Tr\left[(\bar z\mathbb{I}-J^{\dagger})(z\mathbb{I}-J)G\right]G_{ij}}\nn\\
&\frac{c\,\tau}{N}\moy{\Tr[(z\mathbb{I}-J)G][(z\mathbb{I}-J)G]_{ij}}+O(N^{-1})\nn\\
=&-c\,\kappa^2\,R\,\moy{G_{ij}}+c\,\tau\,\bar Q\moy{[(z\mathbb{I}-J)G]_{ij}}+O(N^{-1})\;.
\end{align}
Gathering all the terms we obtain the identity
\begin{align}
\kappa^2\moy{G}&=\mathbb{I}-(\bar z\mathbb{I}-D^{\mu})\moy{(z\mathbb{I}-J)G}+\moy{X^{\dagger}(z\mathbb{I}-J)G}\\
&=\mathbb{I}-c\,\kappa^2\,R\,\moy{G}-((\bar z-c\,\tau\,\bar Q)\mathbb{I}-D^{\mu})\moy{(z\mathbb{I}-J)G}+O(N^{-1})\nn\\
&=\mathbb{I}-c\,\kappa^2\,R\,\moy{G}-((\bar z-c\,\tau\,\bar Q)\mathbb{I}-D^{\mu})((z-c\,\tau\, Q)\mathbb{I}-D^{\mu})\frac{\moy{G}}{1+c\,R}+O(N^{-1})\nn\;.
\end{align}
We can then express $\moy{G}$ explicitly as
\be
\moy{G}=(1+c\,R)\left(\kappa^2(1+c\,R)^2 \mathbb{I}+((\bar z-c\,\tau\,\bar Q)\mathbb{I}-D^{\mu})( (z-c\,\tau\,Q)\mathbb{I}-D^{\mu})\right)^{-1}\;.
\ee
In the limit $\kappa\to 0$, we define the following functions
\begin{align}
&\lim_{\kappa\to 0}c\,\kappa\,R(z,\bar z;\kappa)=\gamma_{c,\tau}(z,\bar z)\;,\\
&\lim_{\kappa\to 0}c\,Q(z,\bar z;\kappa)=\Omega_{c,\tau}(z,\bar z)\;,
\end{align}
that have a finite limit. Using the definition of $R=\moy{\Tr\,G}/N$ and equations (\ref{re_Q}-\ref{im_Q}) , we obtain the identities
\begin{align}
\gamma_{c,\tau}&=\int d\lambda\,\frac{c\,\gamma_{c,\tau}\, n_{\mu}(\lambda)}{\gamma_{c,\tau}^2+\left(y-\tau\,\Im[\Omega_{c,\tau}]\right)^2+\left(x-\tau\,\Re[\Omega_{c,\tau}]-\lambda\right)^2}\;,\label{gamma_eq_1}\\
\Omega_{c,\tau}&=\int d\lambda\,\frac{c\,n_{\mu}(\lambda)}{\gamma_{c,\tau}^2+\left(y-\tau\,\Im[\Omega_{c,\tau}]\right)^2+\left(x-\tau\,\Re[\Omega_{c,\tau}]-\lambda\right)^2}\left(\frac{x-\lambda}{1+\tau}-\frac{i\,y}{1-\tau}\right)\;.
\end{align}
By definition of $G(z,\bar z;\kappa)$, we can obtain an alternative expression of $\gamma_{c,\tau}$ that reads
\be
\gamma_{c,\tau}=\lim_{\kappa\to 0}c\,\kappa\,\int d^2 w \,\frac{\rho(w,\bar w)}{|z-w|^2+\kappa^2}\geq 0\;.
\ee
Using the identity $\lim_{\kappa\to 0} \kappa/(a^2+\kappa^2)=\delta(|a|)$, it can be further rewritten as
\be
\gamma_{c,\tau}=c\,\int d^2 w \,\rho(z+w,\bar z+\bar w)\delta(|w|)\geq 0\;.
\ee
This equation clearly reveals that the trivial solution $\gamma_{c,\tau}=0$ in Eq. \eqref{gamma_eq_1} corresponds to a position $z$ where the density is zero while the non-trivial solution $\gamma_{c,\tau}>0$ corresponds to a position where the density is non-zero. Supposing the former solution is valid, we obtain the system of equations
\begin{align}
\Re[\Omega_{c,\tau}]&=\int d\lambda\,\frac{c\,n_{\mu}(\lambda)\,(x-\tau\,\Re[\Omega_{c,\tau}]-\lambda)}{\left(y-\tau\,\Im[\Omega_{c,\tau}]\right)^2+\left(x-\tau\,\Re[\Omega_{c,\tau}]-\lambda\right)^2}\\
&=\Re\left[\int d\lambda\,\frac{c\,n_{\mu}(\lambda)}{z-\tau\,\Omega_{c,\tau}-\lambda}\right]\;,\nn\\
\Im[\Omega_{c,\tau}]&=-\int d\lambda\,\frac{c\,n_{\mu}(\lambda)\,(y-\tau\,\Im[\Omega_{c,\tau}])}{\left(y-\tau\,\Im[\Omega_{c,\tau}]\right)^2+\left(x-\tau\,\Re[\Omega_{c,\tau}]-\lambda\right)^2}\,\\
&=-\Im\left[\int d\lambda\,\frac{c\,n_{\mu}(\lambda)}{z-\tau\,\Omega_{c,\tau}-\lambda}\right]\;.\nn
\end{align}
Supposing instead the latter solution is valid, one can show the simple identity
\be
\Im\left[\Omega_{c,\tau}\right]=-\frac{y}{1-\tau}\;.
\ee
Using these results, we obtain the final equations
\begin{align}
\gamma_{c,\tau}&=\int d\lambda\,\frac{c\,\gamma_{c,\tau}\,n_{\mu}(\lambda)}{\displaystyle \gamma_{c,\tau}^2+\frac{y^2}{(1-\tau)^2}+\left(x-\Gamma_{c,\tau}-\lambda\right)^2}\;,\label{eq_gamma_app}\\
\Gamma_{c,\tau}=\tau\, \Re\left[\Omega_{c,\tau}\right]&=\int d\lambda\,\frac{c\,\tau\,n_{\mu}(\lambda)\,(x-\lambda-\Gamma_{c,\tau})}{\displaystyle \gamma_{c,\tau}^2+\frac{y^2}{(1-\tau)^2}+\left(x-\Gamma_{c,\tau}-\lambda\right)^2}\;,\label{eq_Gamma_app}
\end{align}
which can be easily seen to be exactly equivalent to the pair \eqref{spb} obtained in the previous section by a different method, after identification
\be
\Gamma_{c,\tau}:=x-q\;,\;\;\gamma_{c,\tau}:=r\;.
\ee

\subsection{Analysis of the mean density}

For $\tau\to 1$ and taking $y=0$, one recovers that $\Gamma_{c,1}\equiv c\,\Re[r(\lambda)]$ is the real part of the resolvent and $\gamma_{c,1}\equiv c\,\Im[r(\lambda)]$ is the imaginary part, where the resolvent satisfies the Pastur equation \cite{P72}
\be
r(\lambda)=\int d\lambda'\,\frac{n(\lambda')}{\lambda-\lambda'-c\,r(\lambda)}\;.\label{pastur_eq}
\ee
In the opposite limit $\tau\to 0$, one has that $\Gamma_{c,\tau}\to 0$ but $\Gamma_{c,\tau}/\tau\to \Gamma_c=O(1)$ while $\gamma_{c,\tau}\to \gamma_{c}=O(1)$ that satisfy
\be
\gamma_c=\int d\lambda\,\frac{c\,\gamma_c\,n_{\mu}(\lambda)}{\gamma_c^2+|z-\lambda|^2}\;,\;\;\Gamma_c=\int d\lambda\,\frac{c\,n_{\mu}(\lambda)(x-\lambda)}{\gamma_c^2+|z-\lambda|^2}\;.
\ee
recovering the results of \cite{K96}.

Let us now use Eqs. (\ref{eq_gamma_app}-\ref{eq_Gamma_app}) to obtain the expression of the average density of $J$. To this purpose, let us use Eq. \eqref{dphi_z} to show that
\begin{align}
\rho(z,\bar z)=&\frac{1}{\pi}\partial_{\bar z}\lim_{\kappa\to 0}Q(z,\bar z;\kappa)=\frac{1}{\pi\,c}\partial_{\bar z} \Omega_{c,\tau}(z,\bar z)\;.\label{rho_omega}
\end{align}
The expression of $\Omega_{c,\tau}(z,\bar z)$ is quite different in the regime where $\gamma_{c,\tau}=0$, where it can be identified with
\be
\Omega_{c,\tau}(z,\bar z)=\int d\lambda\,\frac{c\,n_{\mu}(\lambda)}{z-\tau\,\Omega_{c,\tau}-\lambda}\;,\label{bar_om_eq}
\ee
and in the regime where $\gamma_{c,\tau}>0$, where it reads
\be
\Omega_{c,\tau}(z,\bar z)=\frac{\Gamma_{c,\tau}(z,\bar z)}{\tau}-i\frac{y}{1-\tau}\;,\;\;\gamma_{c,\tau}(z,\bar z)> 0\;.\label{om_gam_y}
\ee
In particular, in the regime where $\gamma_{c,\tau}=0$, it is simple to show that $\Omega_{c,\tau}(z,\bar z)\equiv\Omega_{c,\tau}(z)$ is independent of $\bar z$ such that the density in Eq. \eqref{rho_omega} is zero in this regime. The edge of the support of the density is thus given by the intersection between the regimes $\gamma_{c,\tau}>0$ and $\gamma_{c,\tau}=0$. In particular, the points at the edge of the support are of coordinates $z_e(x_e)=x_e+i y_e(x_e)$ and satisfy
\begin{align}
1&=\int d\lambda\,\frac{c\,n_{\mu}(\lambda)}{\displaystyle\frac{y_e^2(x_e)}{(1-\tau)^2}+\left(x-\Gamma_{c,\tau}(x_e)-\lambda\right)^2}\;,\label{eq_y_e_1}\\
\Gamma_{c,\tau}(x_e)&=\int d\lambda\,\frac{c\,\tau\,n_{\mu}(\lambda)\,(x_e-\lambda-\Gamma_{c,\tau}(x_e))}{\displaystyle\frac{y_e^2(x_e)}{(1-\tau)^2}+\left(x_e-\Gamma_{c,\tau}(x_e)-\lambda\right)^2}\;.\label{eq_y_e_2}
\end{align}
There exists a maximum value $x_e=\lambda_+$ (resp. a minimum value $x_e=\lambda_-$) such that $y_e(\lambda_{\pm})=0$ and the equation above is only valid for $\lambda_-\leq x_e\leq \lambda_+$.

In the regime $\gamma_{c,\tau}(z,\bar z)>0$, we need to compute explicitly the functions $\partial_x\Gamma_{c,\tau}$ and $\partial_{y}\Gamma_{c,\tau}$ to obtain the expression of the density. To this end, we take the derivatives with respect to $x$ and $y$ in equations (\ref{eq_gamma_app}-\ref{eq_Gamma_app}) in the range where $\gamma_{c,\tau}>0$.
Defining the integrals
\be
J_{p,q}(z,\bar z)=\int d\lambda\,\frac{c\,n_{\mu}(\lambda)\,(x-\Gamma_{c,\tau}-\lambda)^p}{\displaystyle \left[\gamma_{c,\tau}^2+\frac{y^2}{(1-\tau)^2}+\left(x-\Gamma_{c,\tau}-\lambda\right)^2\right]^q}\;,\label{Jpq}
\ee
it yields the system
\begin{align}
0=&J_{0,2}\partial_x \gamma_{c,\tau}^2 +2J_{1,2}(1-\partial_x\Gamma_{c,\tau})\,,\\
0=&J_{0,2}\partial_y \gamma_{c,\tau}^2-2J_{1,2}\partial_y\Gamma_{c,\tau}+J_{0,2}\frac{2y}{(1-\tau)^2}\,,\\
\partial_x \Gamma_{c,\tau}=&\tau(1-\partial_x \Gamma_{c,\tau})-\tau J_{1,2}\partial_x \gamma_{c,\tau}^2-2\tau J_{2,2}(1-\partial_x \Gamma_{c,\tau})\,,\\
\partial_y \Gamma_{c,\tau}=&-\tau \partial_y \Gamma_{c,\tau}-\tau J_{1,2}\partial_y \gamma_{c,\tau}^2+2\tau J_{2,2}\partial_y \Gamma_{c,\tau}-2\tau J_{1,2}\frac{y}{(1-\tau)^2}
\end{align}
Solving this system explicitly, one obtains
\begin{align}
&\partial_x \gamma_{c,\tau}^2=-\frac{2J_{1,2}}{(1+\tau(1-2J_{2,2}))J_{0,2}+2\tau J_{1,2}^2}\;,\;\;\partial_y \gamma_{c,\tau}^2=-\frac{2y}{(1-\tau)^2}\label{gamma_d_x}\\
&\partial_x \Gamma_{c,\tau}=1-\frac{J_{0,2}}{(1+\tau(1-2J_{2,2}))J_{0,2}+2\tau J_{1,2}^2}\;,\;\;\partial_y \Gamma_{c,\tau}=0\;.\label{Gamma_d_x}
\end{align}
In particular, we obtain that in the regime where $\gamma_{c,\tau}>0$ one has $\Gamma_{c,\tau}(z,\bar z)\equiv\Gamma_{c,\tau}(x)$, where $x=\Re[z]$ and
\be
\partial_y\left[\gamma_{c,\tau}^2+\frac{y^2}{(1-\tau)^2}\right]=0\;,\label{dy_gam}
\ee
such that similarly $J_{p,q}(z,\bar z)\equiv J_{p,q}(x)$, where the function $J_{p,q}$ is defined in Eq. \eqref{Jpq}. Clearly, one has that $J_{0,2}(x)>0$ and $J_{2,2}(x)>0$ while $J_{1,2}(x)^2\geq 0$, yielding that within the support of the density
\be
0<\partial_x \Gamma_{c,\tau}(x)<1\;,
\ee
which proves Eq. \eqref{phi_double_prime} of the main text.
Inserting the result of Eqs. \eqref{om_gam_y} and \eqref{Gamma_d_x} into Eq. \eqref{rho_omega}
\begin{align}
\rho(z,\bar z)&=\frac{1}{2\pi c}\left[\frac{(\partial_x-i\partial_y)}{\tau}\Gamma_{c,\tau}+\frac{1}{1-\tau}\right]\Theta(\gamma_{c,\tau}(z,\bar z))\;,\label{dens_app}\\
&=\frac{1}{2\pi c}\left[\frac{1}{\tau(1-\tau)}-\frac{J_{0,2}(x)}{\tau(1+\tau(1-2J_{2,2}(x)))J_{0,2}(x)+2\tau^2 J_{1,2}^2(x)}\right]\Theta(\gamma_{c,\tau}(z,\bar z))\;,\nn
\end{align}
where $z=x+i y$ and we use the convention $\Theta(x)=1$ for $x>0$ and $0$ otherwise is the Heaviside step-function. Inside the support of the density, namely for $\gamma_{c,\tau}(z,\bar z)>0$, the density is completely independent of $y=\Im[z]$ as observed in \cite{Tetal21} for $\tau=0$. To connect this expression to Eq. \eqref{6}, one can use that
\be
L_{l,m}(x)=\int \frac{\lambda^l\,n_{\mu}(\lambda)\,d\lambda}{[t(x)+(q(x)-\lambda)^2]^m}=\int \frac{\lambda^l\,n_{\mu}(\lambda)\,d\lambda}{[\gamma_{c,\tau}^2(x,x)+(x-\Gamma_{c,\tau}(x)-\lambda)^2]^m}\;,
\ee
such that
\begin{align}
J_{0,2}(x)&=L_{0,2}(x)\;,\\
J_{1,2}(x)&=L_{0,2}(x)(x-\Gamma_{c,\tau}(x))-L_{1,2}(x)\;,\\
J_{2,2}(x)&=L_{0,2}(x)(x-\Gamma_{c,\tau}(x))^2-2L_{1,2}(x)(x-\Gamma_{c,\tau}(x))+L_{2,2}(x)\;,
\end{align}
from which one can easily check that
\be
D(x)=L_{2,2}(x)-\frac{L_{1,2}(x)^2}{L_{0,2}(x)}=\frac{J_{1,2}(x)^2}{J_{0,2}(x)}-J_{2,2}(x)\;.
\ee
Finally, using this identification one can check that Eq. \eqref{dens_app} does match with \eqref{6}.

In the special case $\tau=0$, we recover the result of \cite{K96}
\be
\rho(z,\bar z)=\frac{1}{\pi c}\left[1+\frac{J_{1,2}^2(x)}{J_{0,2}(x)}-J_{2,2}(x)\right]\;.
\ee
We introduce the density integrated along the $y$ direction
\be
\rho(x)=\int d^{2}z\,\rho(z,\bar z)\delta\left(x-\frac{z+\bar z}{2}\right)=2y_e(x)\rho(x,x)\;.\label{densint_y}
\ee
The function $y_e(x)\geq 0$ can be conveniently re-expressed by integrating Eq. \eqref{dy_gam} between $y=0$ and $y=y_e(x)$, yielding
\be
\gamma_{c,\tau}(x,x)=\frac{y_e(x)}{(1-\tau)}\;.
\ee
Thus, using that $\gamma_{c,1}(x,x)=c\,\Im[r(x)]$, where $r(x)$ satisfies the Pastur equation \eqref{pastur_eq}, the density defined in Eq. \eqref{densint_y} is simply given in the limit $\tau\to 1$ by
\be
\rho(x)=\frac{\gamma_{c,1}(x,x)}{\pi\,c}=\frac{\Im[r(x)]}{\pi}\;.
\ee

\section{Some properties of the electrostatic potential}\label{prop_elec_pot}

We will now use the results derived in the previous sections to derive two important relations (\ref{phi_c}-\ref{phi_tau}) that play an essential role in the derivation of our main results.

Let us now consider some properties of the electrostatic potential
\begin{align}
\Phi(z,\bar z;c,\tau)&=\frac{1}{2}\int d^2 w \, \rho(w,\bar w)\ln|(z-w)(\bar z-\bar w)|\;.\nn
\end{align}
We are particularly interested in the expressions of its derivatives with respect to the parameters $\tau$ and $c$. While the expression for the electrostatic potential itself is not so simple \eqref{Phi_el1}, we have seen in the previous sections that within the support of the density, its derivative with respect to $x=\Re[z]$ has a simple expression, namely
\be
\partial_x \Phi(z,\bar z;c,\tau)=\frac{\Gamma_{c,\tau}(x)}{c\,\tau}\;,
\ee
where $\Gamma_{c,\tau}(x)$ satisfies the system of equations (\ref{eq_gamma_app}-\ref{eq_Gamma_app}). Let us now take derivatives with respect to $c$ and $\tau$ respectively of these two equations. It yields a simple system of equations that can be solved explicitly and reads
\begin{align}
\partial_c \gamma_{c,\tau}^2&=\frac{1+\tau+2J_{1,2}\Gamma_{c,\tau}-2\tau J_{2,2}}{c\left[(1+\tau(1-2J_{2,2}))J_{0,2}+2\tau J_{1,2}^2\right]}\;,\\
\partial_c \Gamma_{c,\tau}&=\frac{J_{0,2}\Gamma_{c,\tau}-\tau J_{1,2}}{c\left[(1+\tau(1-2J_{2,2}))J_{0,2}+2\tau J_{1,2}^2\right]}\;,\\
\partial_\tau \gamma_{c,\tau}^2&=\frac{2 J_{1,2}\Gamma_{c,\tau}}{\tau\left[(1+\tau(1-2J_{2,2}))J_{0,2}+2\tau J_{1,2}^2\right]}\;,\\
\partial_\tau \Gamma_{c,\tau}&=\frac{J_{0,2}\Gamma_{c,\tau}}{\tau\left[(1+\tau(1-2J_{2,2}))J_{0,2}+2\tau J_{1,2}^2\right]}\;,
\end{align}
where we remind that $J_{p,q}(z,\bar z)\equiv J_{p,q}(x)$ is defined in Eq. \eqref{Jpq}. Using these results together with Eqs. (\ref{gamma_d_x}-\ref{Gamma_d_x}), one can now show that inside the support of the density
\begin{align}
\partial_{x,c}\Phi(z,\bar z;c,\tau)&=\partial_c\left[\frac{\Gamma_{c,\tau}(x)}{c\,\tau}\right]=\frac{1}{2c^2}\partial_x\left(\gamma_{c,\tau}^2-\frac{\Gamma_{c,\tau}^2}{\tau}\right)\;,\label{eq_dc_dx}\\
\partial_{x,\tau}\Phi(z,\bar z;c,\tau)&=\partial_\tau\left[\frac{\Gamma_{c,\tau}(x)}{c\,\tau}\right]=-\frac{\partial_x \Gamma_{c,\tau}^2}{2c\tau^2}\;.
\end{align}
Note that as the value of $z\to z_e$ gets to the edge, the function $\gamma_{c,\tau}^2\to 0$ but one has
\be
\lim_{z\to z_e} \partial_x \gamma_{c,\tau}^2(z,\bar z)=\partial_{x_e} \frac{y_e^2(x_e)}{(1-\tau)^2}=-\frac{2J_{1,2}(x_e)}{(1+\tau(1-2J_{2,2}(x_e)))J_{0,2}(x_e)+2\tau J_{1,2}^2(x_e)}\;.
\ee
Using that $\partial_x [y^2/(1-\tau)^2]=0$ for any point within the support of the density, one can replace the function $\gamma_{c,\tau}^2\to \gamma_{c,\tau}^2+y^2/(1-\tau)^2$ in equation \eqref{eq_dc_dx} to have a function that is continuous and has a continuous derivative at the edge. On the other hand, the function $\Gamma_{c,\tau}(x)$ and its derivative are continuous as $x\to x_e$.

Outside of the support of the density, one can compute similarly
\be
\partial_{z} \Phi(z,\bar z;c,\tau)=\frac{\Omega_{c,\tau}(z)}{c}\;,
\ee
where $\Omega_{c,\tau}(z)$ satisfies \eqref{bar_om_eq}. Proceeding similarly as in the support of the density, we obtain the identities
\begin{align}
\partial_{z}\Omega_{c,\tau}(z)&=-\frac{K_2(z)}{1-\tau K_2(z)}\;,\\
\partial_c \Omega_{c,\tau}( z)&=\frac{1}{1-\tau K_2(z)}\frac{\Omega_{c,\tau}(z)}{c}\;,\\
\partial_\tau \Omega_{c,\tau}(z)&=\frac{K_2(z)}{1-\tau K_2(z)}\bar \Omega_{c,\tau}(z)\;,\\
K_2(z)&=\int d\lambda\, \frac{c\,n_{\mu}(\lambda)}{(\bar z-\tau\, \Omega_{c,\tau}(z)-\lambda)^2}\;.
\end{align}
Thus, outside the support of the density, one obtains that
\begin{align}
\partial_{z,c}\Phi(z,\bar z;c,\tau)&=\partial_c\left[\frac{\Omega_{c,\tau}(z)}{c}\right]=-\frac{\tau\,\partial_{z} \Omega_{c,\tau}^2}{2c^2}\;,\\
\partial_{ z,\tau}\Phi(z,\bar z;c,\tau)&=\partial_c\left[\frac{\Omega_{c,\tau}(z)}{c}\right]=-\frac{\partial_{z} \Omega_{c,\tau}^2}{2c}\;.
\end{align}
A similar identity can be obtained by replacing $z\to \bar z$ and $\Omega_{c,\tau}(z)\to \bar \Omega_{c,\tau}(\bar z)$. This yields
\begin{align}
\partial_{x,c}\Phi(z,\bar z;c,\tau)&=-\frac{\tau\,\partial_{x} \left[\Omega_{c,\tau}^2+\bar \Omega_{c,\tau}^2\right]}{4c^2}=\frac{\tau}{2c^2}\,\partial_{x} \left(\Im[\Omega_{c,\tau}]^2-\Re[\Omega_{c,\tau}]^2\right)\;,\\
\partial_{x,\tau}\Phi(z,\bar z;c,\tau)&=\frac{1}{2c}\,\partial_{x} \left(\Im[\Omega_{c,\tau}]^2-\Re[\Omega_{c,\tau}]^2\right)\;.
\end{align}
We may now compute
\begin{align}
&\int_{-\infty}^{x}\partial_{x,c}\Phi(z,\bar z;c,\tau)=\partial_c\Phi(z,\bar z;c,\tau)-\lim_{\Re[w]\to -\infty}\partial_{c}\Phi(w,\bar w;c,\tau)\label{id_d_c}\\
&=\frac{1}{2c^2}\left(\gamma_{c,\tau}^2(z,\bar z)+\frac{\Im[z]^2}{(1-\tau)^2}-\frac{\Gamma_{c,\tau}^2(\Re[z])}{\tau}\right)\nn\\
&+\frac{\tau}{2c^2}\,\left(\Im[\Omega_{c,\tau}(z_e)]^2-\Re[\Omega_{c,\tau}(z_e)]^2-\frac{y_e(x_e)^2}{(1-\tau)^2}+\frac{\Gamma_{c,\tau}^2(\Re[z_e])}{\tau^2}\right)\nn\\
&-\lim_{\Re[ w]\to -\infty}\frac{\tau}{2c^2}\,\left(\Im[\Omega_{c,\tau}(w)]^2-\Re[\Omega_{c,\tau}(w)]^2\right)\nn\;.
\end{align}
One can simply obtain from Eq. \eqref{bar_om_eq} that $\lim_{\Re[ w]\to -\infty}\Omega_{c,\tau}(w)=0$. On the other hand, we can use that as $\Re[z]\to \infty$,
\begin{align}
\partial_c\Phi(z,\bar z;c,\tau)&=\frac{\partial_c}{2}\int d^2 w \, \rho(w,\bar w)\ln|(z-w)(\bar z-\bar w)|\\
&=\frac{\partial_c}{2}\left[\ln|z|+\ln|\bar z|+O(\Re[z]^{-1})\right]\;,\nn
\end{align}
where we have used that $\int d^2 w \, \rho(w,\bar w)=1$ such that $\lim_{\Re[w]\to -\infty}\partial_{c}\Phi(w,\bar w;c,\tau)=0$. Finally, comparing Eq. \eqref{bar_om_eq} and Eqs. (\ref{eq_y_e_1}-\ref{eq_y_e_2}) one can check that as $z\to z_e$, the function
\be
\lim_{z\to z_e}\Omega_{c,\tau}(z)=\frac{\Gamma_{c,\tau}(x_e)}{\tau}-i\,\frac{y_e(x_e)}{(1-\tau)}\;.
\ee
Thus, Eq. \eqref{id_d_c} simplifies considerably and reads for any point $z$ within the support of the density
\be
\partial_c\Phi(z,\bar z;c,\tau)=\frac{1}{2c^2}\left(\gamma_{c,\tau}^2(z,\bar z)+\frac{\Im[z]^2}{(1-\tau)^2}-\frac{\Gamma_{c,\tau}^2(\Re[z])}{\tau}\right)\;.
\ee
Proceeding similarly for the derivative with respect to $\tau$, the identity reads
\be
\partial_{\tau}\Phi(z,\bar z;c,\tau)=-\frac{\Gamma_{c,\tau}^2(\Re[z])}{2c\,\tau^2}\;.
\ee
Taking the point on the real axis, i.e. $z=\bar z=x$, these equations reproduce Eqs. (\ref{phi_c}-\ref{phi_tau}) in the main text.

\end{document}